\documentclass[12pt]{article}

\usepackage[margin=1in]{geometry}

\usepackage[utf8]{inputenc}
\usepackage[T1]{fontenc}
\usepackage{lmodern}

\usepackage[english]{babel}

\usepackage{setspace}
\usepackage{parskip}

\usepackage{graphicx}
\usepackage{graphics}
\usepackage{float}
\usepackage{placeins}
\usepackage{pdflscape}
\usepackage{afterpage}
\usepackage{pdfpages}

\usepackage{booktabs}
\usepackage{array}
\usepackage{tabularx}
\usepackage{multirow}
\usepackage{longtable}
\usepackage{makecell}

\usepackage[labelsep=period,font=sc,skip=0pt,justification=centering]{caption}
\usepackage{subcaption}
\usepackage{amsmath,amssymb,amsthm}
\usepackage{mathtools}
\usepackage{amscd}
\usepackage{amsfonts}
\usepackage{amsbsy}
\usepackage{bm}
\usepackage{dsfont}
\usepackage{cancel}
\usepackage{mathrsfs}
\usepackage{latexsym}
\usepackage{braket}
\usepackage{array}

\usepackage[shortlabels]{enumitem}

\usepackage{xcolor}
\definecolor{AcceptGreen}{RGB}{46,139,87}
\definecolor{WarningOrange}{RGB}{230,126,34}
\definecolor{DangerRed}{RGB}{192,57,43}
\definecolor{CardinalRed}{cmyk}{0,1,0.65,0.34}
\usepackage[colorlinks,linkcolor=CardinalRed,citecolor=CardinalRed,urlcolor=CardinalRed]{hyperref}
\usepackage{attachfile2}
\usepackage{url}

\usepackage{listings}
\lstdefinelanguage{Stata}{
  morekeywords={if, first, cluster},
  sensitive=true,
}

\usepackage[flushmargin]{footmisc}
\usepackage[round]{natbib}

\usepackage{xargs}
\usepackage{lineno}
\usepackage{CJKutf8} 
\usepackage{epsfig}  

\usepackage{titlesec}
\titlelabel{\thetitle.\ }
\titleformat{\section}
{\normalfont\fontsize{15}{15}\bfseries}{\thesection.}{0.5em}{}

\usepackage{titling}
\usepackage{todonotes}
\makeatletter
\providecommand\@dotsep{5}
\def\listtodoname{List of Todos}
\def\listoftodos{\@starttoc{tdo}\listtodoname}
\makeatother

\makeatletter
\def\maxwidth{\ifdim\Gin@nat@width>\linewidth\linewidth\else\Gin@nat@width\fi}
\def\maxheight{\ifdim\Gin@nat@height>\textheight\textheight\else\Gin@nat@height\fi}
\makeatother
\setkeys{Gin}{width=\maxwidth,height=\maxheight,keepaspectratio}

\graphicspath{{./graphs/}}%

\newcommand{\bland}{\begin{landscape}}
\newcommand{\eland}{\end{landscape}}

\newcommand{\dnum}[1]{\textcolor{black}{#1}} 

\newcommand{\burl}[1]{\textcolor{blue}{\url{#1}}}

\newenvironment{itemize*}%
  {\begin{itemize}\setlength{\itemsep}{0pt}\setlength{\parskip}{0pt}}%
  {\end{itemize}}
\newenvironment{enumerate*}%
  {\begin{enumerate}\setlength{\itemsep}{0pt}\setlength{\parskip}{0pt}}%
  {\end{enumerate}}

\newcommand{\beq}{\begin{equation}}
\newcommand{\eeq}{\end{equation}}

\newcommand*\Bigpar[1]{\left( #1 \right)}

\newcommandx{\deriv}[2][1=x,2=f]{\nabla \, #2 \Bigpar{ #1 } }

\renewcommand{\to}{\rightarrow}

\newtheoremstyle{mystyle}
  {12pt}{12pt}{}{}{\sffamily\bfseries}{.}{0.5em}
  {\thmname{#1}\thmnumber{ #2}\thmnote{ (#3)}}
\theoremstyle{mystyle}

\newenvironment{proof-sketch}{\noindent{\bf Sketch of Proof}\hspace*{1em}}{\qed\bigskip\\}
\newenvironment{proof-idea}{\noindent{\bf Proof Idea}\hspace*{1em}}{\qed\bigskip\\}
\newenvironment{proof-of-lemma}[1][{}]{\noindent{\bf Proof of Lemma {#1}}\hspace*{1em}}{\qed\bigskip\\}
\newenvironment{proof-of-proposition}[1][{}]{\noindent{\bf Proof of Proposition {#1}}\hspace*{1em}}{\qed\bigskip\\}
\newenvironment{proof-of-theorem}[1][{}]{\noindent{\bf Proof of Theorem {#1}}\hspace*{1em}}{\qed\bigskip\\}
\newenvironment{inner-proof}{\noindent{\bf Proof}\hspace{1em}}{$\bigtriangledown$\medskip\\}
\newenvironment{proof-attempt}{\noindent{\bf Proof Attempt}\hspace*{1em}}{\qed\bigskip\\}

\newcolumntype{L}[1]{>{\raggedright\let\newline\\\arraybackslash\hspace{0pt}}m{#1}}
\newcolumntype{C}[1]{>{\centering\let\newline\\\arraybackslash\hspace{0pt}}m{#1}}
\newcolumntype{R}[1]{>{\raggedleft\let\newline\\\arraybackslash\hspace{0pt}}m{#1}}

\graphicspath{{./figs/}}%

\begin{document}
  

\title{\Large\bf Scaling Reproducibility: An AI-Assisted Workflow\\for Large-Scale Replication and Reanalysis%
\thanks{Yiqing Xu, Assistant Professor, Department of Political Science, Stanford University. Email: \url{yiqingxu@stanford.edu}. Leo Yang Yang, Research Assistant Professor, Department of Accountancy, Economics and Finance, School of Business, Hong Kong Baptist University, Kowloon, Hong Kong SAR. Email: \url{leoyang@hkbu.edu.hk}. The authors used Claude Code and ChatGPT as research and writing assistants in preparing this manuscript. All interpretations, conclusions, and any errors remain solely the responsibility of the authors. We thank Mike Alvarez and Gary King for their extremely helpful comments, suggestions, and historical context on the introduction of transparency measures in political science.}
\\\bigskip}
\vspace{1.5em}

\author{Yiqing Xu\\(Stanford)\and Leo Yang Yang\\(HKBU)}

\date{\bigskip
First version: February 17, 2026 \\
This version: May 31, 2026 \\
}

\maketitle
\vspace{1em}
\begin{abstract}\onehalfspacing
\vspace{1em}
\noindent Computational reproducibility is central to scientific credibility, yet verifying published results at scale remains costly. We develop an AI-assisted workflow for automated full-paper replication---retrieving materials, reconstructing environments, executing code, and matching outputs to point estimates reported in regression tables. We define a universe of all empirical and quantitative papers from the three top political science journals (2010--2025) and measure stated data availability using automated extraction. For a stratified sample of 384 studies, we apply the workflow to conduct full-paper replication, totaling \dnum{3,523} empirical models. We find that journal verification requirements, combined with data archiving mandates, drive reproducibility: the share of fully or largely reproducible papers rises from \dnum{20.8\%} before DA-RT adoption to \dnum{82.5\%} after, and conditional on accessible replication packages, \dnum{92.1\%} of papers are fully or largely reproducible (234/254). As a secondary application, we apply standardized IV diagnostics to \dnum{84} studies (\dnum{597} IV specifications among \dnum{1,910} replicated models), illustrating how automated execution enables systematic reanalysis across heterogeneous empirical settings.

\bigskip\noindent\textbf{Keywords:} reproducibility, replication, research transparency, open science, AI-assisted workflows, agentic AI, causal inference

\end{abstract}

\thispagestyle{empty}  
\clearpage
\newpage
\doublespacing

\clearpage

\setcounter{page}{1}
\abovedisplayskip=5pt
\belowdisplayskip=5pt
\doublespacing

\section{Introduction}

Reproducibility is fundamental to research credibility and cumulative scientific progress. In empirical social science, reproducible analyses allow researchers to verify published claims, scrutinize identifying assumptions, and assess the practical relevance of new methodological developments. Access to real-world data and code has therefore become important both for assessing research credibility and for advancing methodology through systematic reanalysis.

Institutional norms have expanded the availability of replication packages---bundles of data, code, and documentation deposited by authors to enable reproduction of published results. Leading journals in political science and economics now require authors to post replication packages, and some conduct replication checks before publication. Yet availability alone does not ensure reproducibility at scale. Replication packages vary widely in environments, structure, and execution logic, so reproducing results across many papers remains costly and fragile. The bottleneck is operational: executing heterogeneous packages in a standardized and auditable manner requires substantial effort.

This paper develops an AI-assisted workflow for automated full-paper replication to address this execution bottleneck. The workflow reconstructs computational environments, executes code, and verifies published results. Its purpose is to determine whether existing empirical analyses can be reproduced reliably and at scale, not to introduce new estimators or diagnostics. Once data and code are standardized through this process, they can be readily reused, lowering the cost for methodological researchers to develop and apply new methods on a common empirical basis.

A central design principle is the separation of scientific reasoning from computational execution. Researchers specify the empirical targets to be reproduced---model specifications and reported quantities of interest as defined by the paper's narrative and replication package. Conditional on these inputs, replication reduces to execution-oriented tasks: acquiring replication packages, reconstructing environments, locating and running specifications, extracting analysis datasets, and harmonizing outputs. Throughout this paper, we use \emph{reproducibility} to mean re-execution of existing code and data \citep{nasem2019reproducibility}, distinguishing it from \emph{replicability}---obtaining similar findings in new studies. Earlier literature sometimes uses ``replication'' for the former \citep{king1995replication}.\footnote{The terminology varies across fields. In political science, \citet{king1995replication} used ``replication'' to encompass both re-execution of existing analyses and new studies testing the same hypothesis. Economics has a similar tradition \citep{vilhuber2020reproducibility}. We follow the distinction codified by the National Academies \citep{nasem2019reproducibility} and adopted by recent work in both economics \citep{vilhuber2020reproducibility} and political science \citep{alvarez2022reproduce}: \emph{reproducibility} denotes re-execution of the original code and data, while \emph{replicability} denotes obtaining consistent findings with new data. The compound ``replication package''---the standard term for the bundle of code and data deposited by authors---retains the older usage; the property we measure when re-executing those packages is reproducibility.}

Using this workflow, we study a large set of top political science publications. We define the universe as all empirical and quantitative papers in the \emph{American Political Science Review} (APSR), \emph{American Journal of Political Science} (AJPS), and \emph{Journal of Politics} (JOP) from 2010 to 2025, totaling \dnum{3,464} papers. For this universe, we measure stated data availability using automated extraction of availability statements. Combined with replication outcomes, this allows us to compare trends in stated availability and actual reproducibility over time and relate them to changes in journal policies, particularly the introduction of verification requirements.

We then apply the workflow to two tasks. First, we conduct full-paper replication for a stratified sample of 384 studies (8 papers per journal-year). For each paper, we execute the complete codebase and attempt to reproduce point estimates reported in regression tables, excluding figures, descriptive tables, and simulation results. This defines a consistent unit of evaluation across heterogeneous designs and software environments. The estimated population-level rate of fully or largely reproducible papers rises from \dnum{20.8\%} before DA-RT adoption to \dnum{82.5\%} after, using journal-specific implementation dates (APSR: 2016; AJPS and JOP: 2015). This increase is driven by journal policies requiring public data archiving and, especially, in-house or third-party verification. Conditional on accessible materials, \dnum{234} of \dnum{254} papers (\dnum{92.1\%}) are fully or largely reproducible regardless of publication year, of which \dnum{193} (\dnum{76.0\%}) are fully reproducible---meaning every reported coefficient is matched exactly---and the remaining \dnum{41} are largely reproducible (more than \dnum{80\%} but fewer than 100\% of coefficients matched). Failures trace almost entirely to missing or restricted data, not to code quality.

Second, we apply the workflow to instrumental variable (IV) designs. Building on \citet{lal2024much}, we analyze \dnum{84} studies; the workflow replicates all main-text regressions in each paper (\dnum{1,910} models in total) and extracts \dnum{597} verified IV specifications for standardized diagnostics, illustrating how automated execution enables systematic analysis across heterogeneous empirical settings. The \dnum{84}-study corpus combines \dnum{66} of the \dnum{67} studies in \citet{lal2024much}, plus \dnum{18} newly incorporated studies (2023--2025) under identical inclusion criteria. The one Lal et al.\ paper absent from our corpus is excluded because the Extractor agent produced zero coefficients in the main text, leaving no targets for downstream matching. Conditional on accessible materials, the workflow reproduces benchmark estimates exactly and completes all diagnostic tests.

This level of reliability is achieved through an adaptive, version-controlled process. Replication packages span multiple programming languages, estimation commands, directory structures, and coding conventions, and many failure modes arise only when new materials are encountered. Recurring issues are encoded as generalized rules in the execution layer and versioned across runs. Coverage expands across versions, while numerical behavior remains fixed within each version. The Supplementary Materials document these failure patterns and the corresponding adjustments.

In political science, \citet{king1995replication} argued that scholarly norms should require sufficient data and code to reproduce published results. This vision was formalized through the Data Access and Research Transparency (DA-RT) initiative and the 2014 Journal Editors’ Transparency Statement (JETS), building on public data repositories---principally the Harvard Dataverse \citep{king2007introduction}, ICPSR, and the Open Science Framework---that provided the infrastructure for deposit and retrieval. The three journals in our corpus adopted verification requirements at different times \citep{JETS2014, key2016data}, providing institutional variation that allows us to assess whether these policies are associated with improved reproducibility in practice.

These transparency norms made possible a series of large-scale reanalysis projects \citep{hainmueller2019much, lal2024much, chiu2023causal}, in which much of the effort was devoted to reconstructing environments and harmonizing replication packages rather than developing new methods. Those experiences suggest that, once empirical targets are specified, most remaining work is procedural and therefore automatable. The present workflow operationalizes this insight.

More broadly, this project connects to long-standing concerns about research credibility, from Leamer’s call to ``take the con out of econometrics'' \citep{leamer1983let} to the credibility revolution \citep{angrist2010credibility, torreblanca2026credibility} and the replication crisis in psychology \citep{open2015estimating}. A recent Nature collection further evaluates reproducibility, replicability, and analytical robustness across the social and behavioural sciences at scale \citep{brodeur2026reproducibility, miske2026reproducibility, tyner2026investigating, aczel2026robustness}. These discussions emphasize verifiable analysis and expose the consequences of fragile research pipelines.

The main contribution of this paper is twofold. First, we make computational reproducibility measurable at scale and document that journal transparency policies---particularly verification requirements---are associated with substantially higher reproducibility rates in political science. Second, we show that automated execution enables standardized diagnostics across published studies, illustrated with IV designs, and can facilitate methodological research by providing reusable data and outputs.

The scope of this study is empirical social science, with data drawn from top political science journals where norms for sharing replication packages are relatively strong. The workflow requires usable code and data and does not recover results when packages are missing or fundamentally flawed. The findings therefore reflect what is feasible under current best practices. Although the analysis focuses on political science, the framework is general and can be extended to other fields with established transparency norms.

\section{AI-Assisted Reproducibility Workflow}
\label{sec:workflow}

This section describes the AI-assisted workflow used to conduct full-paper replication and design-specific diagnostics. The workflow targets the execution bottleneck identified in the introduction: executing heterogeneous replication packages in a standardized and auditable manner. It does not automate methodological reasoning or introduce new statistical procedures. Instead, it standardizes and accelerates execution when usable data and code are available.

\paragraph*{Design principles.} 

Large-scale reproducibility involves a basic tension between heterogeneity and determinacy. Replication materials vary widely across studies in programming language (Stata, R, and Python), directory structure, naming conventions, and documentation quality. At the same time, reproducibility requires determinacy: for a fixed pipeline version and fixed inputs, numerical outputs must not depend on ad hoc decisions, platform-specific defaults, or stochastic behavior.

The workflow resolves this tension through three separation principles. First, deterministic orchestration is separated from adaptive coordination. A LangGraph state machine routes tasks across pipeline stages along explicit conditional edges, while an \emph{Adaptive Coordinator}---an LLM-driven sub-agent implemented through Claude Code---is dispatched only when a stage fails its validation gate: it interprets the failure context, identifies the root cause, and applies a minimal repair before control returns to the graph. All numerical work---data preparation, model estimation, and diagnostic computation---is executed by version-controlled program code. For a fixed pipeline version and fixed inputs, the workflow produces identical numerical outputs and retains a complete audit trail. Second, scientific reasoning is separated from execution. Researchers define the empirical targets and, where applicable, the diagnostic templates. The workflow handles the operational tasks: acquiring replication packages, reconstructing environments, executing code, and harmonizing outputs. Third, execution is separated from verification. Running an author’s code successfully does not by itself establish that the paper’s reported results have been reproduced. The workflow therefore treats code execution and result verification as distinct phases, with an explicit gate between them.

\paragraph*{Architecture.} The system is organized as a three-layer architecture. At the top layer, a LangGraph state machine routes tasks between pipeline stages along explicit conditional edges; LLM-driven coordination enters only when a validation gate detects a failure, in which case the Adaptive Coordinator is dispatched to diagnose and apply minimal repairs before control returns to the graph. The middle layer consists of stage-specific agents---most are deterministic Python; the Profiler and Extractor invoke language models for PDF-level metadata and coefficient extraction. At the bottom layer, statistical scripts in R, Stata, and Python perform all estimation and diagnostic computation. The orchestrator controls task routing but is excluded from numerical estimation and inference. Figure~\ref{fig:architecture} illustrates the architecture, using distinct color families---blue with solid arrows for deterministic components, peach with dotted arrows for adaptive ones---to make the boundary between reproducible routing and LLM-driven repair visible at a glance.

Execution proceeds through three phases. In Phase~A (acquisition and execution), the system extracts metadata and replication links from the paper’s PDF, downloads the replication package from public repositories, prepares the code for automated execution by resolving path dependencies and environment assumptions, and executes the complete codebase. All estimation commands encountered during execution are instrumented to capture structured output, including coefficients, standard errors, sample sizes, and clustering. The output is a comprehensive log of every regression the code produces.

\afterpage{%
  \clearpage
  \newgeometry{
    top=2cm,
    bottom=2cm,
    left=2cm,
    right=2cm
  }
\begin{landscape}
\vspace{-2em}\begin{figure}[p]
\begin{center}
\includegraphics[width=1.3\textwidth]{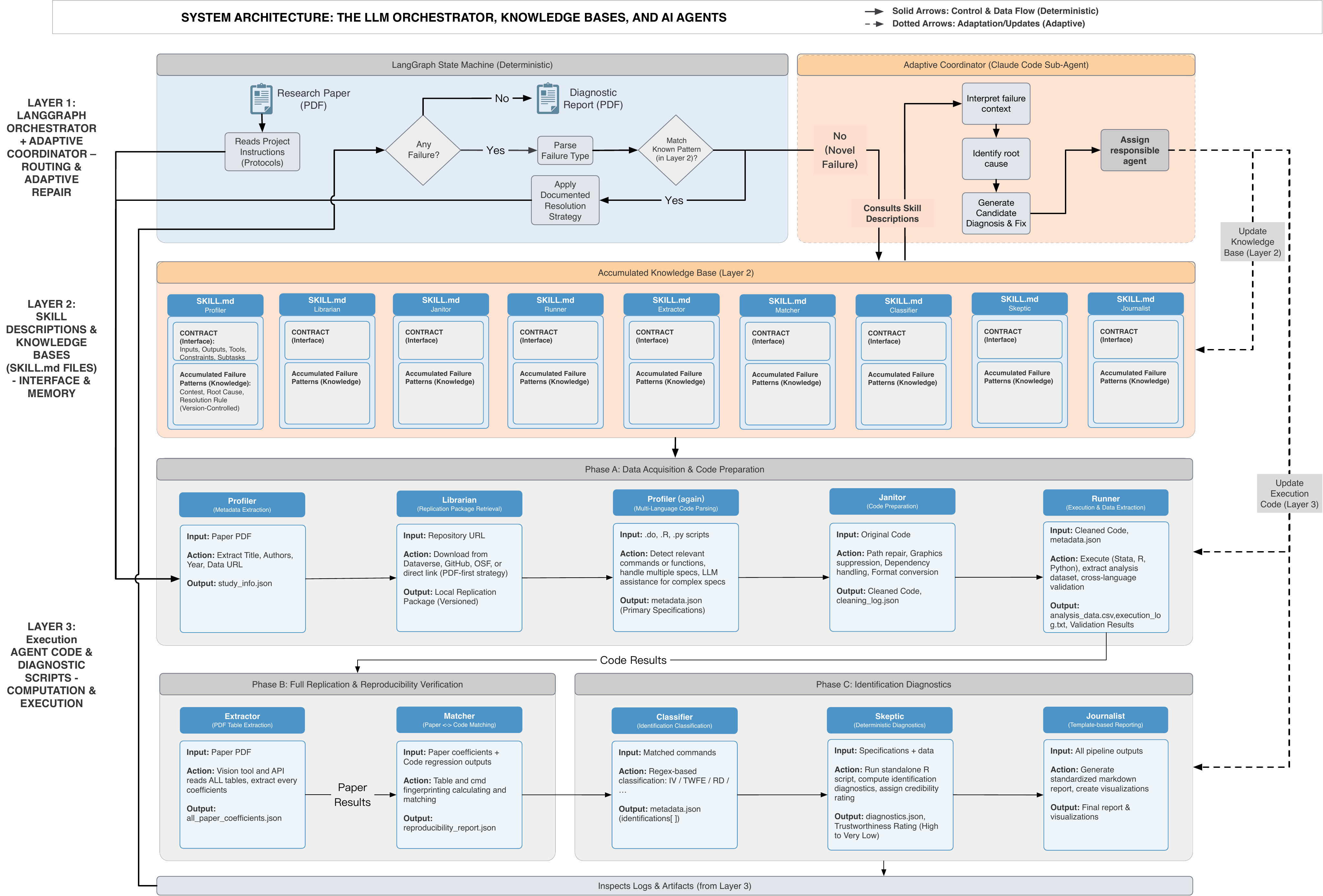}\\ \bigskip
\caption{\textbf{Overview of the AI-assisted workflow for reproducibility.}
{\footnotesize The figure summarizes the system’s three-layer architecture. Blue, solid elements are deterministic components whose outputs are reproducible for fixed inputs. Peach, dotted elements are adaptive components that handle novel failures. Layer~1 routes tasks through a LangGraph state machine. Layer~2 stores \texttt{SKILL.md} files, which define agent contracts and record failure-recovery knowledge. Layer~3 contains stage agents and statistical scripts in R, Stata, and Python. The workflow moves through acquisition and execution (Phase~A), reproducibility verification (Phase~B), and diagnostic evaluation (Phase~C), with gates between phases.
}}
\label{fig:architecture}
\end{center}
\end{figure}
\end{landscape}
\restoregeometry
\clearpage
}

In Phase~B (reproducibility verification), the system extracts all coefficients reported in the paper’s regression tables and compares them against the code-generated outputs. Matching uses precision-aware rounding: each code coefficient is rounded to the same number of decimal places as the paper-reported value, then checked for exact equality. Optimal assignment across paper--code coefficient pairs ensures that each reported value is matched to the most appropriate code output. Phase~B concludes with a reproducibility verdict: papers classified as fully or largely reproducible proceed to Phase~C; others enter a bounded fix loop (up to three iterations) before the verdict is finalized.

In Phase~C (diagnostic evaluation), the system identifies the research design used in each reproduced specification and applies the corresponding diagnostic template. The current implementation includes a diagnostic template for instrumental-variable designs. For IV designs, this includes first-stage $F$-statistics, Anderson--Rubin tests, bootstrap confidence intervals, jackknife sensitivity, and OLS comparisons, implemented through the \texttt{ivDiag} package \citep{ivDiag2024}. Phase~C concludes with a verification check confirming that all diagnostics completed without error. The system then compiles a standardized report combining the reproducibility assessment from Phase~B and the diagnostic results from Phase~C.

Each phase produces explicit intermediate artifacts---structured logs, extracted datasets, matching reports, diagnostic outputs---written to disk. Agents share no hidden state. This modularity ensures that every step is inspectable, and execution can resume from any phase without restarting the pipeline.

\paragraph*{Adaptation.} The workflow was developed iteratively through repeated encounters with diverse replication packages. Replication materials span multiple programming languages, estimation commands, directory structures, and coding conventions, and many failure modes arise only when new materials are encountered. When a recurring failure pattern is identified, the resolution is codified deterministically at one of two levels: \emph{within a stage agent}, by adding a Python rule that handles the specific case during execution, or \emph{within the LangGraph routing layer}, by adding a conditional edge or gate router that dispatches the failure pattern to the appropriate recovery node. The Adaptive Coordinator often proposes such updates during its repair attempts, but each is reviewed and version-controlled before being merged. Over time, this two-level codification reduces the workflow's reliance on adaptive coordination: failure patterns that initially required the Adaptive Coordinator's judgment become deterministic graph edges or agent-level rules in subsequent versions. Coverage expands across pipeline versions, while numerical behavior remains fixed within each version. The executor is kept design-agnostic: extending to new research designs requires specifying a new diagnostic template rather than modifying the underlying architecture.


\section{Empirical Corpus and Evaluation Design}
\label{sec:corpus}

We apply the workflow described in Section~\ref{sec:workflow} to publications from three leading political science journals. The analysis proceeds at three levels. First, we define a universe of empirical and quantitative papers and measure stated data availability at the population level. Second, we draw a stratified sample and attempt full-paper replication. Third, we apply a fixed diagnostic template to a corpus of IV designs. Each level narrows the scope and increases the cost per paper. The first requires only text classification; the second requires executing replication packages; the third requires additional design-specific diagnostics. Table~\ref{tab:design} summarizes the two evaluation tasks.

\begin{table}[!htbp]
\centering\onehalfspacing
\caption{Empirical Task Summary}\label{tab:design}
\resizebox{0.95\textwidth}{!}{
\begin{tabular}{lcc}
\hline\hline\small
 & \textbf{Full-Paper Replication} & \textbf{IV Diagnostics} \\
\hline
Journals & APSR, AJPS, JOP & APSR, AJPS, JOP \\
Period & 2010--2025 & 2010--2025 \\
Universe & 3,464 empirical, quant papers & 66 + 18 IV studies \\
Sample & 384 (8 per journal-year) & 84 papers \\
Models replicated & 3,523 & 1,910 (\textit{597 IV specs}) \\
Coefficients & 18,185 & 5,804 \\
Sampling design & Stratified, inverse-probability-weighted & \citet{lal2024much} + forward extension \\
Evaluation unit & All main-text models & All replicated main-text IV specs \\
Matching criterion & Matching point estimates & Exact 2SLS point estimate \\
Verdict categories & Fully / largely / & Reproduced / not reproduced \\
 & partially / not reproducible & \\
Benchmark verification & Human review of failures & Human review of failures \\
\hline\hline
\end{tabular}}
\begin{minipage}{0.93\textwidth}
\vspace{4pt}
{\footnotesize
\textit{Notes.} The 84 IV studies undergo full-table replication to enable per-specification IV verification; four papers appear in both corpora. Across both corpora the workflow covers \dnum{464} unique papers, \dnum{5,367} replicated models, and \dnum{23,828} coefficients (deduplicating the 4 overlap papers on the IV side). The IV corpus combines 66 of the 67 studies in \citet{lal2024much} (one absent because the Extractor produced zero coefficients in the main text), plus 18 additional studies (2023--2025) using identical inclusion criteria.}
\end{minipage}
\end{table}

\subsection{Policy Context}

The three journals adopted transparency requirements at different times and through different mechanisms. The key distinction is between \emph{availability} policies, which require authors to deposit data and code in a public repository, and \emph{verification} policies, which require that deposited materials be executed and checked before publication.

The discipline-wide catalyst was the 2014 JETS, organized through the DA-RT initiative. Twenty-seven political science journals---including all three in our corpus---committed to requiring that ``cited data are available at the time of publication through a trusted digital repository'' by January 15, 2016. JETS formalized a norm; implementation varied by journal.

AJPS moved earliest. It introduced a mandatory data archiving policy in 2012 and announced a formal replication policy in March 2015 \citep{AJPSReplicationGuide2015}. That same year, AJPS contracted with the Odum Institute for third-party verification of accepted manuscripts---one of the earliest systematic external verification programs in the social sciences. Under this policy, corresponding authors of accepted manuscripts must upload replication files to the AJPS Dataverse, and publication is contingent on successful verification. AJPS thus combined availability and verification before the JETS deadline.

JOP transitioned during 2015, coinciding with a publisher change to the University of Chicago Press and an editorial transition. By mid-2015, published JOP articles carried a standardized statement that replication packages were available in the JOP Dataverse---a shift from earlier practice, when authors were asked only to note the location of packages without mandatory deposit \citep{key2016data}. JOP adopted in-house verification around early 2021, when published articles began stating that the analysis had been ``successfully replicated by the JOP replication analyst.''

APSR signed JETS in 2014 and revised its submission guidelines effective January 2016, but enforcement accelerated under a new editorial team beginning in August 2016. Authors of conditionally accepted manuscripts must now submit a reproducibility package to the APSR Dataverse, and the journal reviews packages to reproduce tables and figures prior to publication. 

These policy shifts define three regimes: (i)~pre-policy, when data sharing was voluntary or weakly enforced; (ii)~post-availability, when journals required data deposit but did not verify reproducibility; and (iii)~post-verification, when journals required that submitted code and data reproduce the published results as a condition for acceptance. The timing of each transition differs by journal.

\subsection{Defining the Population}

We accessed articles published in the APSR, AJPS, and JOP between 2010 and 2025. %
Each article was classified by an LLM as empirical and quantitative or not, based on full text, following the criteria in \citet{torreblanca2026credibility}.\footnote{For this task, we used Claude Sonnet 4 Batch API; model identifier \texttt{claude-sonnet-4-20250514}.} This classification identifies papers whose primary analysis is based on quantitative data from real-world sources, excluding purely theoretical, qualitative, or simulation-based studies. After applying the current screened political-science inclusion filter, the resulting universe contains \dnum{3,464} papers: \dnum{902} from APSR, \dnum{955} from AJPS, and \dnum{1,607} from JOP.

For each paper in the universe, we extract the data availability statement from the full text and record two indicators: (i)~whether the paper includes a stated commitment to make replication packages available, and (ii)~whether the statement provides a link to a public repository where the packages can be located. These indicators measure stated availability at the population level. Whether the packages can actually be downloaded and executed is assessed only for the replication sample described below.

\subsection{Replication Sample and Reproducibility Criterion}

From the universe of \dnum{3,464} empirical papers, we draw a stratified sample of \dnum{384} studies: \dnum{8} papers per journal-year across \dnum{48} cells. Because the number of empirical papers varies across journal-years, we apply inverse probability weights to ensure that sample-level estimates reflect the composition of the universe.

For each sampled paper, we execute the authors' complete codebase and compare every regression coefficient it produces against every coefficient reported in the paper's regression tables. Point estimates are the most precisely defined and consistently reported quantities across heterogeneous studies; exact coefficient agreement implies that the full specification---controls, fixed effects, sample restrictions, and data transformations---was correctly reproduced. We exclude descriptive tables, simulation results, and figures: descriptive tables lack the standardized structure of regression output, and figures would require computer vision to extract numerical targets. Standard errors and other uncertainty measures are captured for downstream diagnostics but are not used as matching criteria, as they are reported inconsistently across papers and can depend on bootstrap seeds. Each code-generated coefficient is rounded to the same number of decimal places as the paper-reported value, then checked for exact equality. Code-generated and paper-reported coefficients are then matched via optimal one-to-one assignment.

We classify a paper as \emph{fully reproducible} if 100\% of point estimates in main-text regression tables are matched, \emph{largely reproducible} if more than 80\% but fewer than 100\% are matched, \emph{partially reproducible} if 50--80\% are matched, and \emph{not reproducible} otherwise. The match rate is computed at the coefficient level---the share of individual point estimates that reproduce exactly---rather than at the column or table level. This finer granularity preserves credit for columns in which most coefficients reproduce but one or two do not, while still requiring a perfect match for the strict ``fully reproducible'' verdict. Because we use the more demanding coefficient-level criterion, the ``fully reproducible'' category in this study is correspondingly conservative; we therefore report ``fully or largely reproducible'' as the primary headline measure.

\subsection{IV Diagnostic Corpus}

As a secondary application, we apply the same workflow to a corpus of IV designs. The starting point is the 67 IV studies analyzed in \citet{lal2024much}, drawn from the same three journals (2010--2022) and satisfying a common set of design restrictions: linear IV models with a single endogenous regressor. The original project manually verified benchmark 2SLS point estimates for each study, which serve as fixed ground truth.

We extend this corpus by incorporating 18 additional IV studies published between 2023 and 2025 under identical inclusion criteria, yielding a working corpus of 84 studies (66 from the original Lal et al.\ (2024) sample plus 18 newly incorporated). One paper from the Lal et al.\ (2024) sample is absent because the Extractor agent produced zero paper coefficients, leaving no targets for downstream matching. For each study, the workflow replicates all main-text IV specifications, including baseline and robustness variants. The total number of evaluated specifications increases from \dnum{70} in the original Lal et al.\ (2024) corpus to \dnum{597} in the extended corpus.

The IV evaluation targets the IV specifications reported in each paper's main-text regression tables that successfully pass replication---a departure from prior practice, including our own earlier version of this workflow, of randomly sampling a fixed number of IV specifications per paper for diagnostic analysis. Authors frequently report dozens or hundreds of IV regressions in supplementary appendices and robustness sections, but the specifications they elevate to the main-text tables are the ones they themselves regard as substantively important. Focusing on this set produces a diagnostic corpus that reflects the IV evidence underlying each paper's central claims rather than a representative sample of the authors' robustness exercises. Each candidate specification is verified through a three-step chain: (i)~the authors' code reproduces the paper's coefficients, (ii)~an independent R re-estimation matches the code output, and (iii)~only fully verified specifications proceed to diagnostic analysis. For the 18 newly incorporated studies, we further verify that the automated pipeline reproduces the reported 2SLS estimate.

\section{Demonstration}
\label{sec:demonstration}

We trace a single study through the full workflow to illustrate both tasks described in Section~\ref{sec:corpus}. We select \cite{Chong2019}, which examines whether a randomized get-out-the-vote (GOTV) campaign in rural Paraguay affects female electoral participation. The paper exemplifies an experimental IV design with a clean over-identified instrument set: three indicators for assignment to door-to-door treatment at three intensities (\texttt{D2D30}, \texttt{D2D40}, \texttt{D2D50}) serve as instruments for the actual local treatment proportion (\texttt{ratio\_treat}), with voting in the 2013 presidential election (\texttt{elecc\_presid2013}) as the outcome. The replication code is written in Stata---the dominant environment in our evaluation corpus---and the workflow processes the paper in two steps: full-paper replication and, for the IV specifications, design-specific diagnostics.

\paragraph*{Full-paper replication.} Before applying design-specific diagnostics, the workflow attempts to reproduce all coefficients reported in the paper's regression tables. It executes the complete Stata replication package and compares every code-generated coefficient against the corresponding paper-reported value using precision-aware rounding and optimal assignment. All \dnum{36} coefficients reported in Table~2 of the paper---\dnum{12} from Panel~A (effect on registration), \dnum{12} from Panel~B (effect on turnout), and \dnum{12} from the IV results panel---are reproduced exactly. The paper is classified as fully reproducible.

\paragraph*{IV Diagnostic analysis.} The workflow identifies all six IV specifications reported in Table~2 of \cite{Chong2019}. All share the same outcome (\texttt{elecc\_presid2013}), endogenous treatment (\texttt{ratio\_treat}), instrument set (\texttt{D2D30}, \texttt{D2D40}, \texttt{D2D50}), and clustering variable. The six specifications differ along two dimensions: campaign content (\emph{Targeted} vs \emph{Untargeted} door-to-door messaging) and the modeling assumption on the treatment-density first stage (\emph{Full Sample}, \emph{Linear}, \emph{Nonlinear}).

For each specification, the workflow applies a three-step verification chain before proceeding to diagnostics: (i)~the authors' code reproduces the paper's coefficients, (ii)~an independent R re-estimation matches the code output, and (iii)~only fully verified specifications receive diagnostic analysis. All six specifications in \cite{Chong2019} pass this chain.

\begin{table}[!th]
\centering
\setstretch{1.15}\footnotesize
\caption{Diagnostic results for \cite{Chong2019}: all six main-text IV specifications}
\label{tab:chong_diagnostics}
\resizebox{\textwidth}{!}{%
\begin{tabular}{lcccccc}
\hline\hline
 & Spec 1 & Spec 2 & Spec 3 & Spec 4 & Spec 5 & Spec 6 \\ \hline
Campaign & Targeted & Targeted & Targeted & Untargeted & Untargeted & Untargeted \\
First-stage form & Full Sample & Linear & Nonlinear & Full Sample & Linear & Nonlinear \\
\hline
\multicolumn{7}{l}{\textit{Instrument strength}} \\
\quad Effective $F$ & 48.3 & 30.9 & 25.3 & 37.2 & 20.5 & 25.3 \\
\hline
\multicolumn{7}{l}{\textit{2SLS estimate}} \\
\quad Coefficient & 0.124 & $-$0.040 & 0.231 & 0.079 & $-$0.056 & 0.124 \\
\quad Std.\ error & 0.056 & 0.071 & 0.075 & 0.060 & 0.063 & 0.086 \\
\quad $p$-value & 0.027 & 0.570 & 0.002 & 0.191 & 0.375 & 0.146 \\
\quad $N$ & 3{,}350 & 1{,}122 & 2{,}228 & 3{,}434 & 1{,}154 & 2{,}280 \\
\quad Clusters & 282 & 109 & 173 & 284 & 109 & 175 \\[3pt]
\hline
\multicolumn{7}{l}{\textit{Robust inference}} \\
\quad Anderson--Rubin $p$ & 0.092 & 0.815 & 0.002 & 0.269 & 0.778 & 0.225 \\
\quad Analytic 95\% CI & [0.014, 0.235] & [$-$0.179, 0.098] & [0.084, 0.378] & [$-$0.039, 0.198] & [$-$0.181, 0.068] & [$-$0.043, 0.292] \\
\quad Boot-$c$ 95\% CI & [0.004, 0.236] & [$-$0.243, 0.152] & [0.032, 0.369] & [$-$0.038, 0.215] & [$-$0.289, 0.102] & [$-$0.050, 0.321] \\
\quad Boot-$t$ incl.\ 0? & No & Yes & No & Yes & Yes & Yes \\[3pt]
\hline
\multicolumn{7}{l}{\textit{Sensitivity (jackknife)}} \\
\quad Range & [0.116, 0.131] & [$-$0.060, $-$0.017] & [0.200, 0.248] & [0.070, 0.091] & [$-$0.086, $-$0.030] & [0.092, 0.157] \\
\quad Max $\Delta$\% & 6.9 & 56.5 & 13.2 & 15.3 & 51.7 & 26.1 \\[3pt]
\hline
\multicolumn{7}{l}{\textit{OLS comparison}} \\
\quad OLS coefficient & 0.083 & $-$0.064 & 0.170 & 0.043 & $-$0.102 & 0.130 \\
\quad 2SLS/OLS ratio & 1.49 & 0.63 & 1.36 & 1.85 & 0.55 & 0.96 \\[3pt]
\hline
\end{tabular}}
\begin{minipage}{\textwidth}
\vspace{4pt}
\footnotesize
\textit{\textbf{Notes:}} All six specifications use outcome \texttt{elecc\_presid2013} (voted in 2013 presidential election), endogenous treatment \texttt{ratio\_treat} (local proportion treated), instrument set \{\texttt{D2D30}, \texttt{D2D40}, \texttt{D2D50}\} (door-to-door assignment intensity), and cluster on locality. Bootstrap tests use 1{,}000 iterations with cluster-level resampling. Jackknife removes one cluster at a time. Because the instrument set is over-identified, the Anderson--Rubin test is informative for weak-instrument-robust inference.
\end{minipage}
\end{table}

The diagnostic template, described in Section~\ref{sec:tests} in the Supplementary Materials, is then applied to each verified specification. Table~\ref{tab:chong_diagnostics} reports the results. All six specifications display strong first stages, with effective $F$-statistics ranging from \dnum{20.5} to \dnum{48.3}, well above the conventional cutoff of 10; no specification is flagged for weak instruments. Because the instrument set is over-identified (three instruments for one endogenous regressor), the workflow also reports Anderson--Rubin tests, which deliver weak-instrument-robust inference and agree with the main 2SLS inference across all six specifications. Substantively, the Targeted-Full and Targeted-Nonlinear specifications yield significant positive effects on voting ($p < 0.05$), while the Untargeted subgroups show null effects; the workflow applies the same criteria uniformly and reports these results without substantive interpretation. The reanalysis does not assess the credibility of the core identification assumptions: the random-assignment-as-instrument design makes the exclusion restriction plausible by construction, but the workflow does not adjudicate exclusion or monotonicity claims that lie outside the diagnostic template.

\paragraph*{Standardized report.} The workflow compiles all outputs---replication results, verification chain, and diagnostic analyses---into a standardized report. The report includes an executive summary listing reproducibility verdicts and diagnostic summaries, followed by specification-level diagnostics with coefficient comparison plots, first-stage $F$-statistics, bootstrap confidence intervals, and jackknife sensitivity analyses. The format is identical across papers, facilitating cross-study comparison.

The full pipeline---from PDF ingestion and package retrieval to report generation---ran end-to-end without human intervention. For this paper, wall-clock time was dominated by the IV diagnostic stage (200-replicate cluster bootstrap across six specifications); runtime characteristics for the corpus as a whole are reported in Section~\ref{sec:replication}.

\FloatBarrier


\section{Main Findings}\label{sec:findings}

This section reports three sets of findings from deploying the AI-assisted workflow at scale. First, we document trends in stated data and code availability across the three journals over the study period. Second, we report full-paper replication results for the stratified sample of \dnum{384} papers, including inverse-probability-weighted reproducibility rates and a diagnostic classification of non-reproducible cases. Third, we present results from the IV diagnostic corpus, showing that the automated pipeline reproduces the core patterns documented in \citet{lal2024much}. Across both corpora, the workflow processed \dnum{5,367} targeted models in \dnum{464} unique papers (\dnum{3,523} in the \dnum{384}-paper main corpus, plus \dnum{1,844} additional models from the \dnum{80} non-overlapping IV papers) with no human intervention.

\subsection{Data and Code Availability}\label{sec:availability}

Before examining replication outcomes, we summarize the transparency environment in which the workflow operates. For each paper in the universe of \dnum{3,464} empirical articles, the Profiler agent extracts whether the paper includes a data availability statement and whether a repository URL is provided. These indicators measure \emph{stated} availability at the population level. For the stratified replication sample ($n = \dnum{384}$), we go further and verify whether replication packages are actually downloadable and whether they contain usable data.

\afterpage{%
\begin{landscape}
\vspace{-2em}\begin{figure}[p]
\begin{center}
\includegraphics[width=1.3\textwidth]{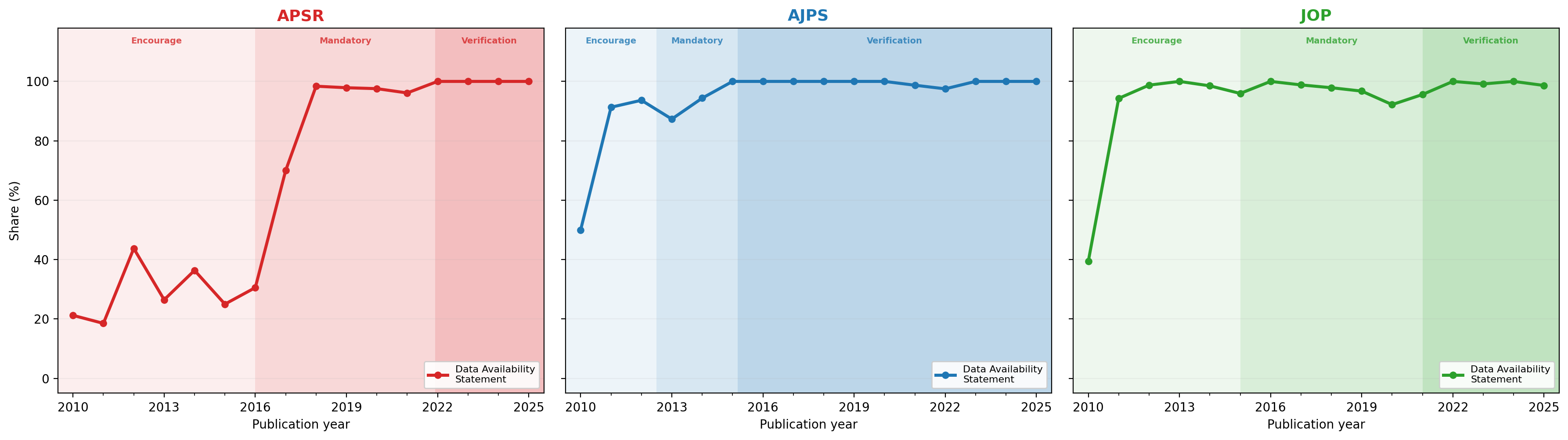}\\[1.5em]
\includegraphics[width=1.3\textwidth]{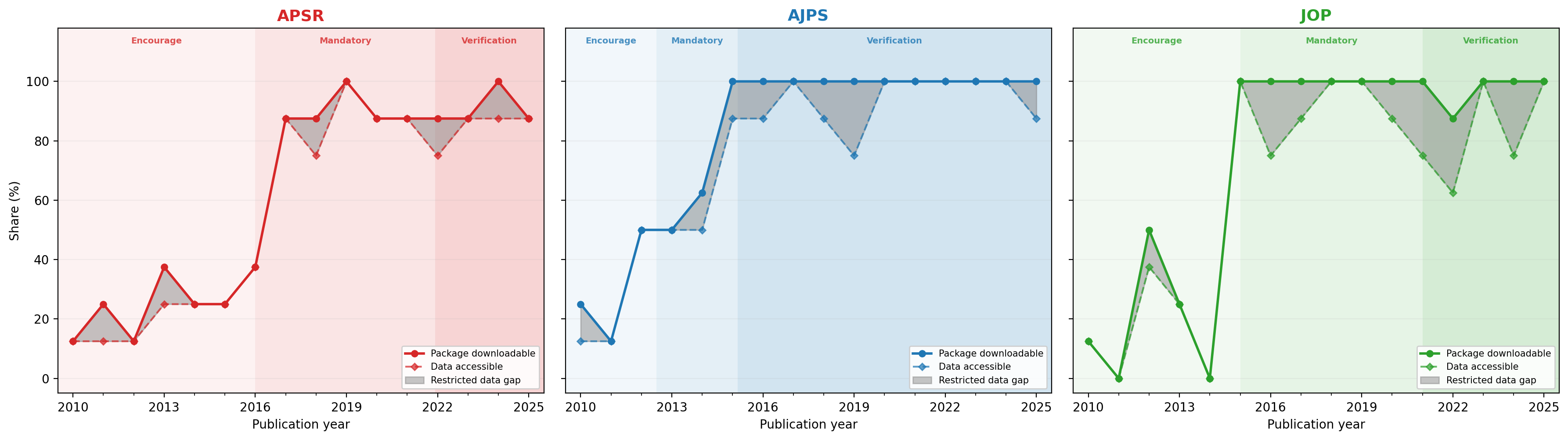}\\[1.5em]
\caption{\textbf{Replication package availability by journal over time.}
{\footnotesize \textit{Upper panel:} Population-level statistics from \dnum{3,464} empirical articles. Each dot represents the share of papers with a stated replication package by journal (APSR, AJPS, JOP) in a given year. Shaded regions indicate policy regimes: pre-policy (no mandatory data archiving), post-availability (mandatory deposit), and post-verification (mandatory reproducibility checks).
\textit{Lower panel:} Sample-level validation from the stratified replication sample ($n = \dnum{384}$). Solid lines with circles show papers whose replication packages are actually downloadable; dashed lines with diamonds show papers with usable (non-restricted) data for at least one model. The gray shaded area between the two lines represents papers with an available replication package but entirely restricted data.}}
\label{fig:availability}
\end{center}
\end{figure}
\end{landscape}
}

Figure~\ref{fig:availability} presents both perspectives. The upper panel plots population-level stated availability by journal. Availability is near zero for all three journals before 2012. For APSR, the sharp increase coincides with the adoption of DA-RT around 2016, with availability climbing steeply after 2019--2020 as the journal adopted in-house verification. For AJPS and JOP, stated availability has been high since the early 2010s. By the end of the study period, stated availability exceeds 90\% in all three journals.

The lower panel provides a more intensive validation using the replication sample. Two lines track progressively stricter definitions of availability: actually downloadable replication packages, and packages containing usable (non-restricted) data. The gap between these lines is revealing. Even for AJPS and JOP, where stated availability appears high early on (upper panel), \emph{consistent} data availability---where downloadable packages reliably contain usable data---only emerges after the adoption of DA-RT around 2015--2016. The shaded area between the two lines captures papers with available replication packages but entirely restricted data. In the post-DA-RT period, restricted data emerges as the primary remaining barrier to reproducibility: authors comply with archiving requirements by posting code and non-restricted data, but key analyses depend on proprietary, confidential, or third-party datasets that cannot be included. Addressing this gap---through secure remote execution, synthetic data, or coordinated access agreements---is an important frontier for the field.

These trajectories reflect a broader institutional transformation. Journal mandates were effective in part because authors could deposit packages in established public repositories---principally the Harvard Dataverse, ICPSR, and the Open Science Framework---that provide persistent access, DOI-based citation, and programmatic retrieval. The infrastructure preceded the mandates and made compliance feasible at scale. The implication for reproducibility is direct: without retrievable replication packages, reproduction cannot begin. Papers published before mandatory archiving are far less likely to have accessible packages, and this shapes the results reported in the next subsection.

\subsection{Full-Paper Replication at Scale}\label{sec:replication}

We now turn to the main empirical task: full-paper replication of \dnum{384} studies sampled from the three journals across 2010--2025. For each paper, the workflow attempts to download the replication package, execute all code, extract every coefficient from the paper’s regression tables, and match code outputs to paper-reported values.

\begin{figure}[!htbp]
\centering
\includegraphics[width=0.85\textwidth]{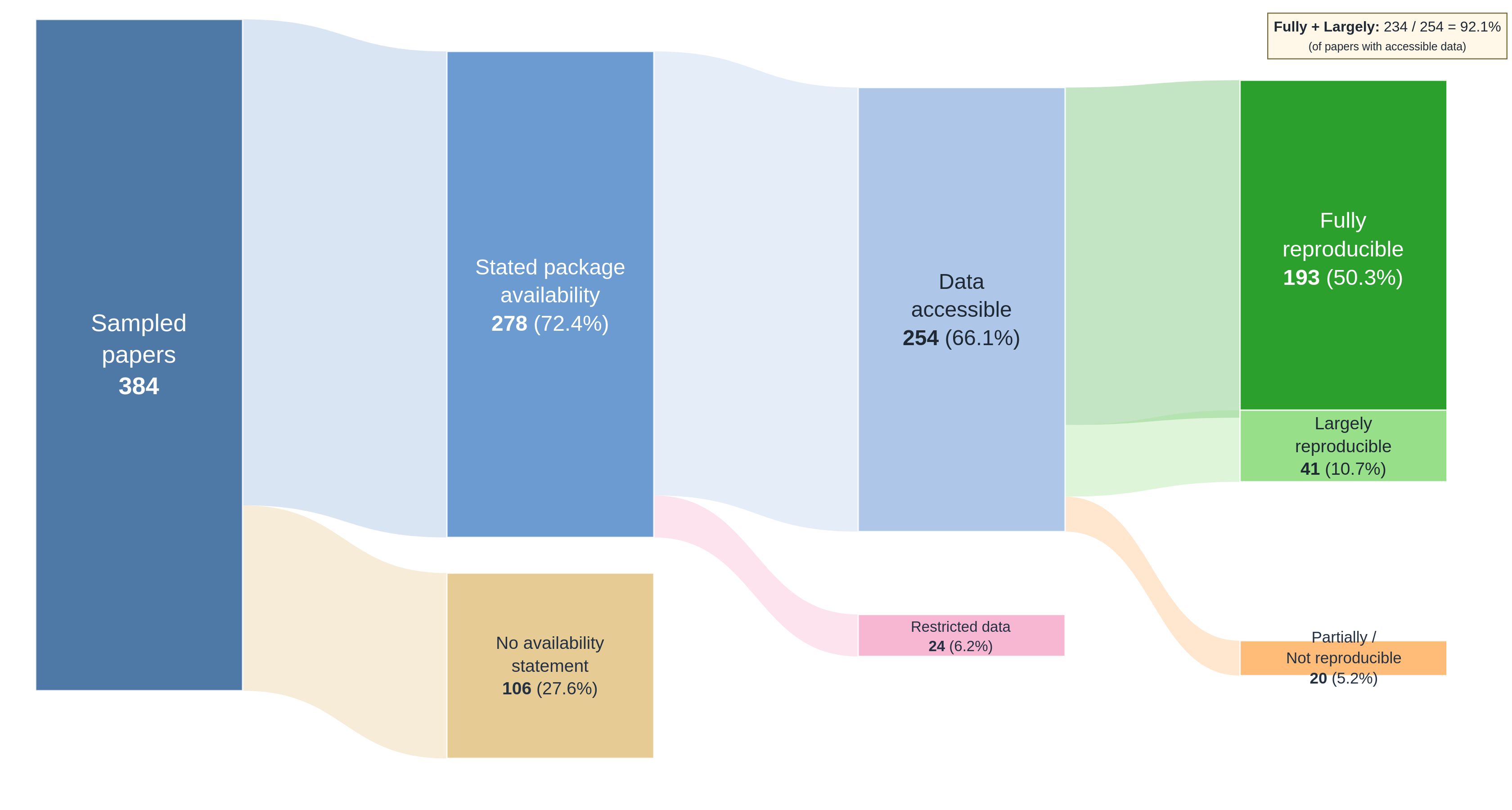}
\caption{\textbf{Replication pipeline funnel.} Flow from \dnum{384} sampled papers to final reproducibility verdicts. Papers exit the pipeline at three stages: no replication package provided (\dnum{106}), restricted or inaccessible data or code (\dnum{24}), and below fully reproducible (\dnum{61}). Among papers with accessible data and executable code, \dnum{234} of \dnum{254} (\dnum{92.1\%}) are fully or largely reproducible, of which \dnum{193} (\dnum{76.0\%}) are fully reproducible under the strict coefficient-level criterion.}
\label{fig:funnel}
\end{figure}

Of the \dnum{384} sampled papers, \dnum{278} (\dnum{72.4\%}) have replication files available in a public repository; the remaining \dnum{106} have no replication package posted. Among papers with available files, \dnum{254} (\dnum{66.1\%} of the full sample) have data that can be accessed and code that can be executed; \dnum{24} papers are blocked by restricted or proprietary data or code. Of the \dnum{254} papers with accessible data and executable code, \dnum{234} (\dnum{92.1\%}) are classified as fully or largely reproducible---\dnum{193} fully reproducible (\dnum{76.0\%}) plus \dnum{41} largely reproducible---and the remaining \dnum{20} are partially or not reproducible (Figure~\ref{fig:funnel}).

The dominant source of non-replication is the absence of replication packages, not code failure: missing packages (\dnum{106}) and restricted data or code (\dnum{24}) together account for \dnum{130} of the \dnum{191} papers that are not fully reproducible.\footnote{That said, our workflow was able to fix repairable coding errors in 10 studies; these are classified as fully reproducible. See Table~\ref{tab:diagnostic-repaired} in the Supplementary Materials.} Among the \dnum{61} papers with accessible materials that fall below the strict fully-reproducible threshold, \dnum{41} are largely reproducible (more than \dnum{80\%} but fewer than \dnum{100\%} of coefficients match), \dnum{10} are partially reproducible, and \dnum{10} are not reproducible---see Table~\ref{tab:diagnostic-gap} in the Supplementary Materials for a detailed breakdown. The adaptive execution layer contributes materially to coverage by repairing common author-side packaging errors that would otherwise block execution, but it cannot overcome packages that were never posted.

\afterpage{%
\begin{landscape}
\vspace{-2.5em}\begin{figure}[p]
\begin{center}
\includegraphics[width=1.3\textwidth]{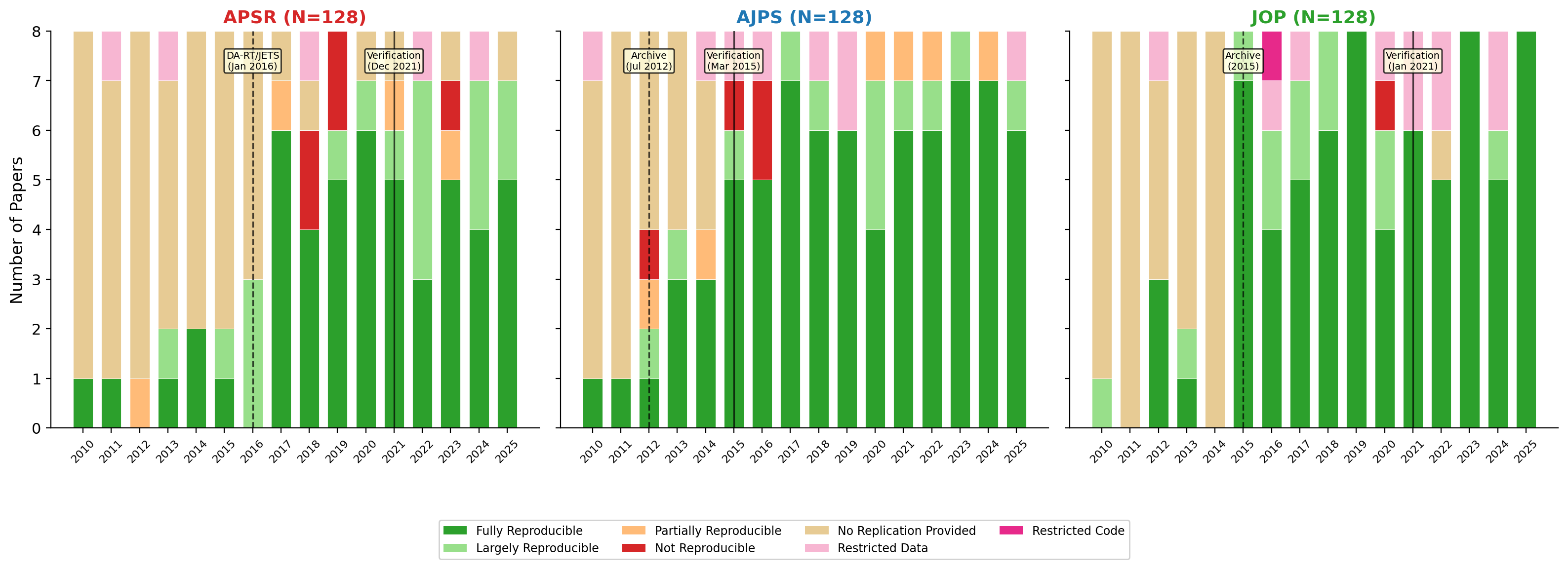}\\[0.5em]
\includegraphics[width=1.3\textwidth]{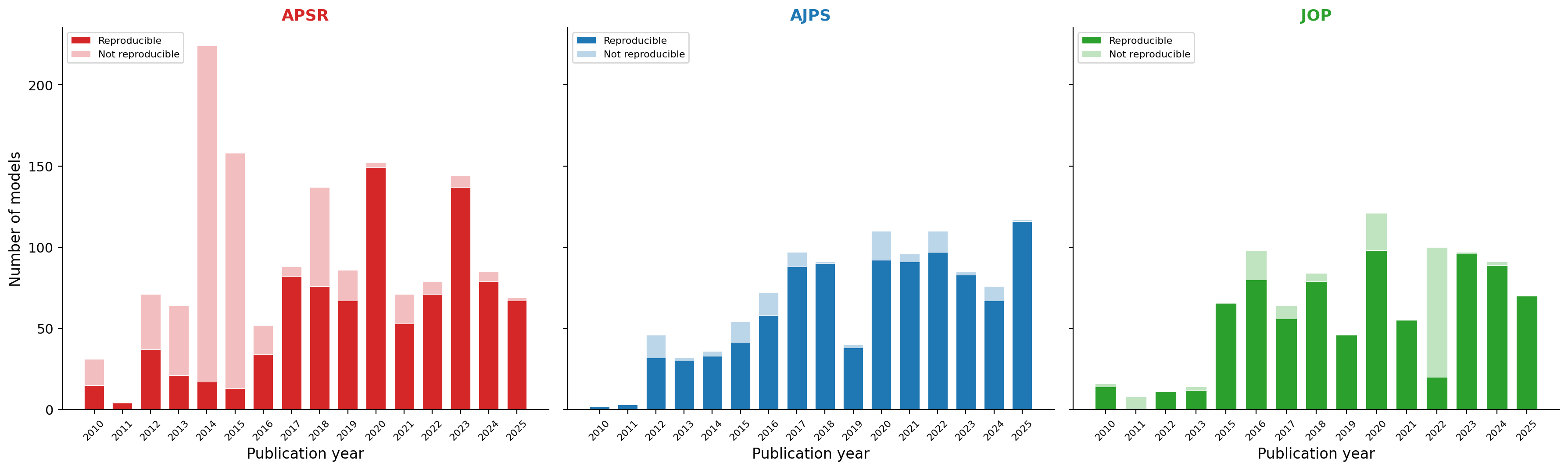}\\[0.5em]
\caption{\textbf{Reproducibility by journal over time.}
{\footnotesize \textit{Upper panel:} Stacked bars show paper-level verdict composition for each journal-year cell ($n = 8$ papers per cell). Green: fully reproducible. Tan: no replication package provided. Other colors: largely, partially, or not reproducible, and restricted data.
\textit{Lower panel:} Model-level reproducibility conditional on data availability. For papers with downloadable, non-restricted replication packages, bars show the number of models attempted in each journal-year cell. Solid shading indicates replicated models; lighter shading indicates non-replicated models.}}
\label{fig:by_year}
\label{fig:model_replic}
\end{center}
\end{figure}
\end{landscape}
}

\paragraph*{Reproducibility by journal and year.} Figure~\ref{fig:by_year} disaggregates these outcomes by journal and publication year. The upper panel shows paper-level verdicts: papers published before DA-RT adoption are dominated by missing replication packages (tan bars), while papers published after are predominantly fully reproducible (green bars). The timing of this shift closely tracks the adoption of mandatory data archiving and verification policies documented in Figure~\ref{fig:availability}, strongly suggesting that institutional policy change---not secular trends in author behavior---drove the improvement. The lower panel shifts from paper-level verdicts to model-level outcomes, restricting attention to papers with downloadable, non-restricted replication packages. For each journal-year cell, bars show the number of models the workflow attempted to replicate, with solid shading for replicated models and lighter shading for non-replicated ones. The contrast between panels is informative: while the upper panel shows substantial non-replication driven by missing data, the lower panel reveals that conditional on data availability, the share of non-reproducible models is very small across all journals and time periods. Mandated verification by a third party or a journal's editorial team, however, does not guarantee full computational reproduction.

\paragraph*{Population-level estimates.} Table~\ref{tab:ipw} quantifies these patterns using inverse-probability-weighted rates, with journal-specific DA-RT implementation dates as cutoffs (APSR: 2016; AJPS and JOP: 2015). Because the number of empirical papers varies across journal-years---from \dnum{27} to \dnum{177}---the inverse-probability-weighting (IPW) estimator reweights each cell to reflect its share of the publication universe. The estimand throughout this table is the population-level share of papers that are fully or largely reproducible. Across all years, the estimated pooled reproducibility rate is \dnum{67.0\%} with a 95\% CI of [\dnum{63.1\%}, \dnum{71.0\%}]. The rate is substantially higher in the post-DA-RT period: \dnum{82.5\%} ([\dnum{77.8\%}, \dnum{87.3\%}]) compared to \dnum{20.8\%} ([\dnum{13.8\%}, \dnum{27.8\%}]) before DA-RT adoption. 
\begin{table}[!htbp]
\centering
\caption{Reproducibility Rates by Journal and DA-RT Period}
\label{tab:ipw}\small
\begin{tabular}{llccc}
\hline\hline
Journal & & All Years & Pre-DA-RT & Post-DA-RT \\
\hline
\multirow{3}{*}{APSR ($\geq$2016)}
 & Rate           & 61.3\% & 16.9\% & 73.2\% \\
 & 95\% CI        & [53.3, 69.3] & [7.3, 26.5] & [63.4, 83.0] \\
 & Papers sampled & 128 & 48 & 80 \\
\hline
\multirow{3}{*}{AJPS ($\geq$2015)}
 & Rate           & 65.7\% & 28.7\% & 84.6\% \\
 & 95\% CI        & [59.0, 72.5] & [15.1, 42.3] & [77.0, 92.1] \\
 & Papers sampled & 128 & 40 & 88 \\
\hline
\multirow{3}{*}{JOP ($\geq$2015)}
 & Rate           & 71.0\% & 15.7\% & 86.8\% \\
 & 95\% CI        & [65.0, 77.0] & [5.1, 26.4] & [79.7, 93.9] \\
 & Papers sampled & 128 & 40 & 88 \\
\hline
\multirow{3}{*}{Pooled}
 & Rate           & 67.0\% & 20.8\% & 82.5\% \\
 & 95\% CI        & [63.1, 71.0] & [13.8, 27.8] & [77.8, 87.3] \\
 & Papers sampled & 384 & 128 & 256 \\
\hline\hline
\end{tabular}
\begin{minipage}{0.95\textwidth}
\vspace{4pt}
{\footnotesize
\textit{Notes.} Each journal-year cell contains $n = 8$ sampled papers. Weights are inverse sampling probabilities: $w_{jt} = N_{jt} / n_{jt}$, where $N_{jt}$ is the number of empirical articles in journal $j$, year $t$. The IPW estimates the share of fully or largely reproducible papers in the publication universe (\emph{fully reproducible}: 100\% coefficient match; \emph{largely reproducible}: more than \dnum{80\%} but fewer than 100\% match). 95\% confidence intervals use the ratio estimator variance (Supplementary Materials, Section~\ref{sec:ipw_formula}). Cluster bootstrap CIs (5{,}000 resamples of journal-year cells) yield consistent intervals (Table~\ref{tab:ipw_bootstrap}). ``Pre-DA-RT'' and ``Post-DA-RT'' use journal-specific cutoffs aligned with each journal’s adoption of mandatory data archiving or verification: APSR $\geq$2016, AJPS $\geq$2015, JOP $\geq$2015.}
\end{minipage}
\end{table}

In the post-DA-RT period, the three journals converge to broadly similar rates: JOP at \dnum{86.8\%}, AJPS at \dnum{84.6\%}, and APSR at \dnum{73.2\%}. AJPS and JOP, both of which introduced DA-RT in 2015 and conduct active verification, cluster tightly around \dnum{85\%}; APSR (\dnum{73.2\%}), which adopted DA-RT in 2016 and ramped up in-house verification only around 2021, continues to lag. Pre-DA-RT rates vary more widely---AJPS at \dnum{28.7\%} versus APSR at \dnum{16.9\%} and JOP at \dnum{15.7\%}---reflecting AJPS’s earlier culture of data sharing even before formal mandates. The timing of the pre-to-post shift closely tracks verification adoption at each journal, consistent with the conclusion that active verification---not merely archiving mandates---drove the improvement in reproducibility.

\paragraph*{Runtime and cost.} The workflow processes the full sample of \dnum{384} papers, and processing is fully parallelizable. For papers where no public replication package is found, the pipeline terminates after profiling and retrieval attempts in under \dnum{2}~minutes. For papers that enter full execution, the median end-to-end time is about \dnum{11}~minutes when the pipeline runs without invoking the Adaptive Coordinator---the \emph{happy path}, taken by \dnum{82\%} of the \dnum{254} executable papers; the \dnum{18\%} requiring Adaptive Coordinator intervention take a median of \dnum{33}~minutes. Anthropic API cost is similarly modest: median \$\dnum{0.54} per paper on the happy path---essentially the Extractor's call to Claude Opus 4.7 for PDF coefficient extraction---and \$\dnum{19}--\$\dnum{56} (interquartile range) when the Adaptive Coordinator is invoked, for a total of approximately \$\dnum{2{,}000} across the executable corpus. By comparison, manual full-paper replication at this scale would require person-years of effort; see the April 2026 reproducibility collection \citep{brodeur2026reproducibility, miske2026reproducibility, tyner2026investigating, aczel2026robustness} for the empirical scope of such efforts.

\subsection{IV Diagnostic Results}\label{sec:iv_results}

As a secondary application, we evaluate the workflow on the IV diagnostic corpus described in Section~\ref{sec:corpus}. This corpus provides a validation anchor: the 67 original studies in \citet{lal2024much} have manually verified benchmark estimates, allowing direct assessment of the workflow’s execution reliability.

For the original 67 papers in \citet{lal2024much}, the workflow autonomously retrieves, executes, and analyzes \dnum{66} (\dnum{98.5}\%). The 1 failure (Blair2022) occurs at the material retrieval stage---the replication archive is no longer publicly available from author-provided sources.\footnote{The data required to reproduce the IV results remain available in the replication materials of \citet{lal2024much}, but we code them as unavailable because our goal is for the model to run the workflow end to end.} The 18 newly incorporated papers (2023--2025) all succeed end to end, yielding a working corpus of \dnum{84} studies with \dnum{597} verified specifications. The high conditional success rate reflects the adaptive nature of the execution layer, the strong compliance with verification requirements among post-DA-RT publications, and the deterministic nature of the diagnostic template.

At the paper level---using the same exact coefficient-match criterion applied to the full-paper sample---\dnum{76} of \dnum{84} retrieved papers (\dnum{90.5\%}) are fully or largely reproducible conditional on retrieval, comparable to the rate in the full-paper sample. The IV diagnostics that follow draw from the \dnum{597} specifications that pass the three-step verification chain described in Section~\ref{sec:corpus}; these include verified specifications from papers whose paper-level verdict falls below the strict fully-reproducible threshold, because the verification chain operates per specification rather than per paper.

On average, this yields \dnum{7.1} verified IV specifications per paper. The largest counts come from papers that report many specifications in the main text, often varying the instrument or sample restriction. This represents more than a twofold increase over the fixed \dnum{3}-per-paper random sample used in the earlier version of this workflow (\dnum{252} versus \dnum{597} specifications). By targeting the full set of main-text IV regressions rather than a random subset, the diagnostic corpus now covers, for each paper, the specifications that the authors themselves elevated to the main text as substantively important rather than an arbitrary slice of their robustness exercises.

\begin{figure}[!ht]
\centering
\includegraphics[width=1\textwidth, trim=0cm 0cm 0cm 2cm, clip]{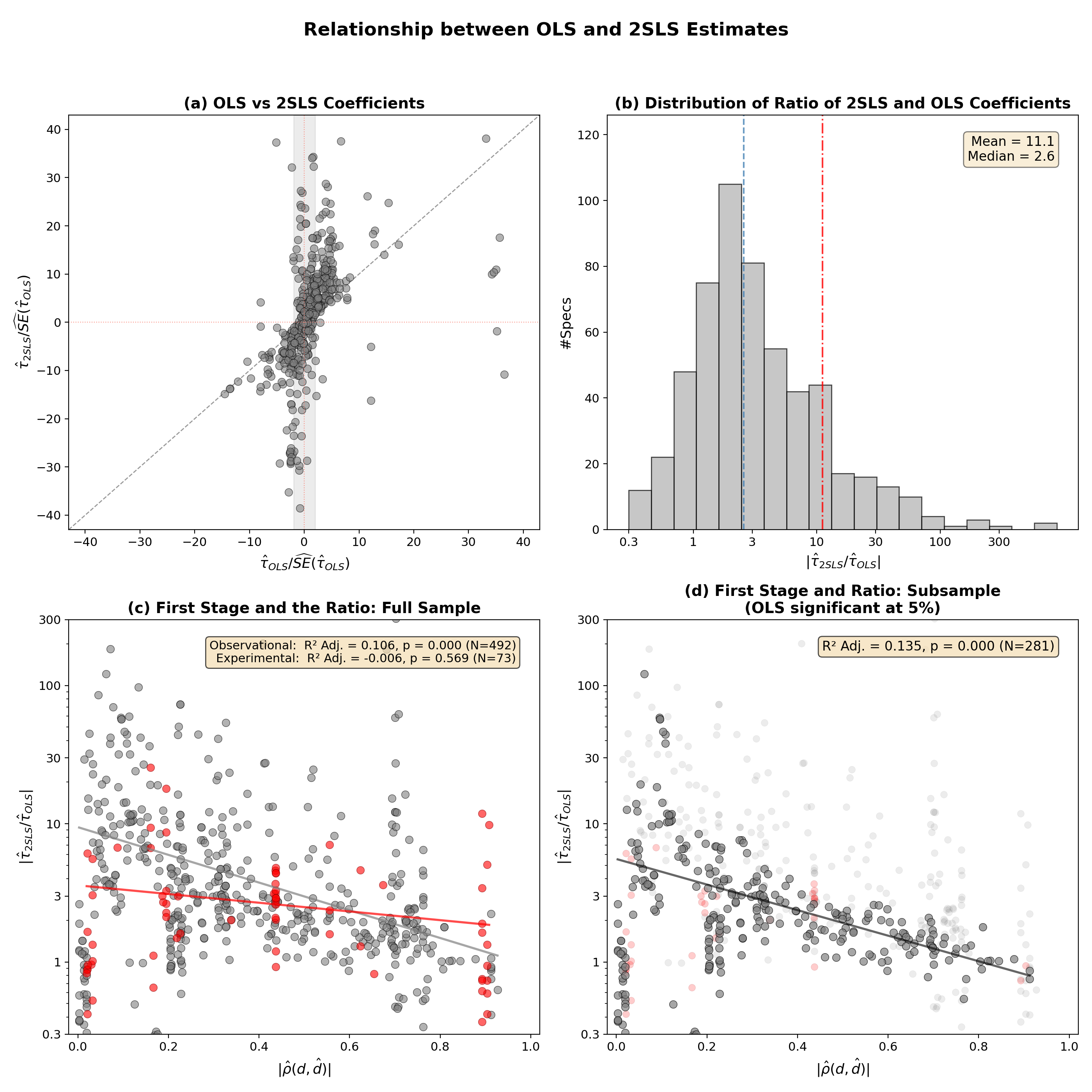}\medskip
\caption{\textbf{Relationship between OLS and 2SLS estimates.}
{\footnotesize This  figure replicates Figure~5 in \citet{lal2024much} using \dnum{597} verified specifications from \dnum{84} papers (versus 70 specifications in the original study). Specifications are drawn from the IV corpus, restricted to those passing per-spec verification (metadata-level diagnosable). Panel~(a) rescales both coefficients by the reported OLS standard errors; the shaded region corresponds to the interval $[-1.96, 1.96]$. Panel~(b) presents the distribution of the log absolute ratio $|\hat{\tau}_{\text{2SLS}} / \hat{\tau}_{\text{OLS}}|$. Panels~(c) and~(d) examine how first-stage strength, measured by $|\hat{\rho}(d,\hat{d})|$, relates to the magnitude of the 2SLS-to-OLS ratio. Gray markers denote observational designs and red markers denote experiment-based instruments.}}
\label{fig:2SLSvOLS}
\end{figure}

\paragraph*{Empirical patterns.} We apply the diagnostic template from \citet{lal2024much} to the extended corpus of \dnum{597} specifications for \dnum{84} studies. Figure~\ref{fig:2SLSvOLS} mirrors Figure~5 in \citet{lal2024much}. The core empirical finding is preserved: 2SLS estimates systematically exceed OLS estimates in magnitude, and the 2SLS-to-OLS ratio is negatively correlated with first-stage strength among observational designs ($p < 0.001$), but not among experimental studies ($p = 0.569$). We cluster the standard errors at the design level, by which we mean each unique outcome-treatment-instrument triplet.

The extended sample confirms the original finding: 2SLS magnitudes systematically exceed OLS (median ratio \dnum{2.57}), and the discrepancy grows as first-stage strength weakens among observational designs, consistent with bias amplification under violations of instrument validity. The pattern is absent among experimental designs. These results reinforce the argument in \citet{lal2024much} that many observational IV estimates rest on fragile identification assumptions. The key difference is scale and speed: whereas the original study required approximately four years of manual data collection, processing, and reanalysis, the automated workflow processed the expanded corpus within days.

\FloatBarrier

\section{Discussion}

This paper shows that computational reproducibility in political science can be verified at scale using an AI-assisted workflow, and that journal verification requirements are associated with substantially higher reproducibility rates. Among \dnum{384} sampled papers from three top journals over 2010--2025, the estimated population-level share of fully or largely reproducible papers rises from \dnum{20.8\%} before DA-RT adoption to \dnum{82.5\%} after. The workflow also enables standardized diagnostics at scale, reproducing and extending the IV analysis of \citet{lal2024much} across \dnum{597} specifications in days rather than years. Together, these results demonstrate that automated verification is feasible, that existing transparency policies have worked, and that the remaining frontier is expanding data access.

These results rest on institutional infrastructure that took decades to build. \citet{king1995replication} argued that scholarly norms should require sufficient data and code to reproduce published results. The DA-RT initiative and the 2014 JETS formalized that argument into journal policy. Public repositories---the Harvard Dataverse \citep{king2007introduction}, ICPSR, the Open Science Framework---provided the deposit and retrieval layer that made compliance feasible. AJPS's contract with the Odum Institute for third-party verification (2015), and the AEA Data Editor system in economics (2019) \citep{vilhuber2020reproducibility}, demonstrated that active verification, not merely archiving, is what drives compliance. Our results show, at scale, that these investments paid off: the pre-to-post DA-RT shift accounts for the bulk of the difference in reproducibility rates. Making data and code publicly available is an important first step; verification infrastructure ensures that posted materials actually work. The present workflow can substantially reduce the cost of providing that verification.

The workflow scales evaluation rather than defining evaluative standards: it relies on diagnostic templates specified by human experts and applies them uniformly across studies. Human-led replication initiatives---journal data editor programs, research transparency organizations such as OSF and BITSS, the Institute for Replication (I4R), and independent replication communities---remain indispensable for evaluating identification strategies, measurement choices, and research design.

All results are conditional on the existence and accessibility of replication packages. The workflow does not take the extra step of fetching data that may be publicly available but not included in the replication package, such as public-use surveys or government datasets hosted on external platforms. When such data are absent from the package, the paper is classified as having inaccessible packages, even though a researcher willing to locate and download the data separately could potentially reproduce the results. The high conditional reproducibility rate (\dnum{92.1\%} fully or largely reproducible among papers with accessible packages, of which \dnum{76.0\%} are fully reproducible at the strict 100\% coefficient threshold) therefore characterizes the quality of self-contained packages, not the reproducibility of the literature as a whole. Similarly, the matching criterion targets point estimates only; extending verification to figures, uncertainty estimates, and other reported quantities is a natural next step. The workflow also does not currently handle studies that are highly computationally intensive, such as large-scale simulation studies and Bayesian analyses requiring extended MCMC sampling. These studies can in principle be incorporated with additional computational resources and are a natural target for future extension.

The workflow can be extended along three dimensions. First, we plan to apply it to additional research designs for which structured corpora already exist, including two-way fixed effects studies \citep{chiu2023causal} and heterogeneous treatment effect analyses using linear interaction models \citep{hainmueller2019much, liu2025practical}. Second, the scope of replication can be deepened beyond benchmark specifications to reconstruct complete tables and figures, verify internal coherence across reported results, and automate diagnostics such as cluster jackknife, leave-one-out influence, and alternative inference procedures. Third, the workflow can integrate into the research and publication pipeline---assisting authors in preparing replication packages at submission, and enabling journals or third-party verification services to conduct reproducibility checks more routinely. A companion study extending this framework across six fields is in preparation.

Beyond these extensions, the broader implications concern how verification, reporting, and methodological development are organized.

\paragraph*{Lowering the cost of verification dramatically.}

The most immediate implication is a substantial reduction in the marginal cost of verification. When replication packages are available, recomputing estimates and applying standardized diagnostics becomes considerably less expensive than under current practice. This does not resolve disputes about identification or theory, but it changes incentives. At present, systematic verification is rare because its cost often exceeds its expected benefit. As that cost falls, more journals may find it feasible to require in-house or third-party reproducibility checks as a condition for acceptance. Authors, anticipating a higher likelihood of auditing, may adopt more disciplined coding practices and address influence, clustering, and resampling concerns ex ante. Verification thus becomes more closely integrated into the publication process rather than applied only after controversies arise.

\paragraph*{Standardizing diagnostic reporting.}

Uniform diagnostic protocols can also reshape reporting norms. If weak-instrument tests, robust inference procedures, and sensitivity analyses are implemented automatically and summarized in standardized formats, discretion in how robustness is selected and presented declines. Journals may require structured diagnostic summaries alongside main results, much as data availability statements have become routine. Referees may increasingly expect influence diagnostics and alternative clustering checks as part of the baseline empirical presentation. Over time, graduate training may adapt to treat such diagnostics as integral components of empirical analysis rather than supplementary exercises.

\paragraph*{Enabling large-scale reanalysis and accelerating methodological research.}

Harmonized analysis datasets and structured metadata enable large-scale reanalysis under consistent criteria. Each of our previous large-scale reanalysis projects---\citet{hainmueller2019much}, \citet{lal2024much}, and \citet{chiu2023causal}---took three to four years of sustained effort. Much of that time was spent harmonizing replication packages, clarifying benchmark estimands, and standardizing robustness checks across heterogeneous applications. With the present workflow, many of these steps can be automated, substantially reducing the time required to conduct comparable large-scale reanalyses.

In the near term, this infrastructure may support more frequent and systematic reassessment of empirical literatures. Research groups, professional associations, or journals could periodically revisit published findings using updated diagnostic standards without incurring multi-year coordination costs. As harmonized corpora accumulate, empirical claims may compete not only on substantive grounds but also on demonstrated stability under shared diagnostics.

A large quantity of harmonized data will have profound implications for methodological research. In computer science, benchmark datasets such as ImageNet \citep{Deng2009}, MS COCO \citep{Lin2014}, SQuAD \citep{Rajpurkar2016}, and GLUE \citep{Wang2018GLUE} structured progress by providing common evaluation environments. Researchers could compare algorithms under identical tasks and metrics, which facilitated cumulative improvement. Analogously, a large collection of harmonized empirical datasets with standardized diagnostic outputs can serve as a benchmark platform for causal and statistical methods. Methodologists could evaluate new estimators and inference procedures across diverse real-world applications rather than relying primarily on stylized simulations. By lowering the cost of empirical validation, the workflow may help shift methodological research toward cumulative comparison under shared empirical settings.
\bigskip

Taken together, these implications suggest that agentic AI workflows for reproducibility, with humans in the loop, can function as research infrastructure. They do not replace researchers' substantive judgment, but make systematic evaluation easier to conduct and harder to avoid. By lowering the cost of verification, standardizing diagnostics, and accelerating methodological development, they may help make transparency and cumulative scrutiny part of routine empirical practice.

\vspace{3em}

\onehalfspacing
\bibliographystyle{apsr}
\bibliography{refs.bib}

\clearpage


\appendix
\onehalfspacing
\setcounter{page}{1}
\setcounter{table}{0}
\setcounter{figure}{0}
\setcounter{equation}{0}
\setcounter{footnote}{0}
\renewcommand\thetable{S\arabic{table}}
\renewcommand\thefigure{S\arabic{figure}}
\renewcommand{\thepage}{S-\arabic{page}}
\renewcommand{\theequation}{A\arabic{equation}}
\renewcommand{\thefootnote}{A\arabic{footnote}}

\vfill
\begin{center}
{\Large\bf Supplementary Materials}\\\medskip{\large\bf Scaling Reproducibility: An AI-Assisted Workflow\\for Large-Scale Replication and Reanalysis}
\end{center}
\vspace{3em}

{\bf
\begin{enumerate}\itemsep1ex
  \item[A.] Additional Tables and Results
  \begin{itemize}
    \item[A.1.] IPW Estimator and Variance Formula
    \item[A.2.] Diagnostic Classification
  \end{itemize}
  \item[B.] System Architecture and Workflow
  \begin{itemize}
    \item[B.1.] Three-Layer Architecture
    \item[B.2.] Agents and Pipeline Stages
    \item[B.3.] Adaptive Execution
  \end{itemize}
  \item[C.] IV Diagnostic Template
  \item[D.] Empirical Inventory and Performance
    \begin{itemize}
    \item[D.1.] Classes of Implementation Variation
    \item[D.2.] Pipeline Performance
    \item[D.3.] Detailed Inventory of Resolved Issues
  \end{itemize}
\end{enumerate}

}
\vfill
\clearpage

\section{Additional Tables and Results}
\label{sec:supplementary_tables}

\subsection{IPW Estimator and Variance Formula}
\label{sec:ipw_formula}

The IPW-weighted reproducibility rate reweights each journal-year cell by its share of the publication universe. Let $j = 1, \ldots, J$ index journals and $t = 1, \ldots, T$ index years. For each cell $(j,t)$, let $N_{jt}$ denote the number of empirical articles in the universe, $n_{jt}$ the number sampled, and $k_{jt}$ the number classified as fully or largely reproducible. Define $\hat{p}_{jt} = k_{jt}/n_{jt}$ and let $N = \sum_{j,t} N_{jt}$. The estimator is
\begin{equation}
\hat{R}
= \frac{\sum_{j,t} N_{jt} \hat{p}_{jt}}{\sum_{j,t} N_{jt}}
= \sum_{j,t} \frac{N_{jt}}{N}\hat{p}_{jt}.
\end{equation}
Equivalently, with inverse probability weights $w_{jt} = N_{jt}/n_{jt}$,
\begin{equation}
\hat{R}
= \frac{\sum_{j,t} w_{jt} k_{jt}}{\sum_{j,t} w_{jt} n_{jt}}.
\end{equation}

Under stratified simple random sampling without replacement within each cell, the variance is
\begin{equation}
\mathrm{Var}(\hat{R})
= \sum_{j,t} \left(\frac{N_{jt}}{N}\right)^2
\left(1 - \frac{n_{jt}}{N_{jt}}\right)\frac{S_{jt}^2}{n_{jt}},
\end{equation}
where $S_{jt}^2$ is the finite-population variance of the binary reproducibility indicator in cell $(j,t)$. For a binary outcome, a plug-in estimator is
\begin{equation}
\widehat{\mathrm{Var}}(\hat{R})
=
\sum_{j,t}
\left(\frac{N_{jt}}{N}\right)^2
\left(1 - \frac{n_{jt}}{N_{jt}}\right)
\frac{\hat{p}_{jt}(1-\hat{p}_{jt})}{n_{jt}-1},
\end{equation}
with standard error $\widehat{\mathrm{SE}}(\hat{R}) = \sqrt{\widehat{\mathrm{Var}}(\hat{R})}$.

Confidence intervals in Table~\ref{tab:ipw} use the Wald interval $\hat{R} \pm 1.96 \times \widehat{\mathrm{SE}}(\hat{R})$, clipped to $[0,1]$. As a robustness check, we also compute cluster bootstrap confidence intervals by resampling journal-year cells with replacement (5{,}000 replicates, percentile method). Table~\ref{tab:ipw_bootstrap} reports the bootstrap intervals; they are similar to the analytic intervals across all subgroups.
\clearpage

\noindent
\begin{minipage}{\textwidth}
\captionsetup{type=table}
\caption{Estimated Reproducibility Rates: Cluster Bootstrap Confidence Intervals}
\label{tab:ipw_bootstrap}\small
\centering
\begin{tabular}{llccc}
\hline\hline
Journal & & All Years & Pre-DA-RT & Post-DA-RT \\
\hline
\multirow{3}{*}{APSR ($\geq$2016)}
 & Rate           & 61.3\% & 16.9\% & 73.2\% \\
 & 95\% CI        & [46.8, 71.9] & [8.4, 23.3] & [64.2, 81.0] \\
 & Papers sampled & 128 & 48 & 80 \\
\hline
\multirow{3}{*}{AJPS ($\geq$2015)}
 & Rate           & 65.7\% & 28.7\% & 84.6\% \\
 & 95\% CI        & [50.6, 79.4] & [15.4, 41.9] & [79.4, 89.4] \\
 & Papers sampled & 128 & 40 & 88 \\
\hline
\multirow{3}{*}{JOP ($\geq$2015)}
 & Rate           & 71.0\% & 15.7\% & 86.8\% \\
 & 95\% CI        & [54.2, 84.2] & [2.6, 28.3] & [78.9, 94.7] \\
 & Papers sampled & 128 & 40 & 88 \\
\hline
\multirow{3}{*}{Pooled}
 & Rate           & 67.0\% & 20.8\% & 82.5\% \\
 & 95\% CI        & [57.4, 75.1] & [12.7, 29.0] & [77.6, 87.5] \\
 & Papers sampled & 384 & 128 & 256 \\
\hline\hline
\end{tabular}
\begin{minipage}{0.92\textwidth}
\vspace{4pt}
{\footnotesize
\textit{Notes.} Point estimates are identical to Table~\ref{tab:ipw}. Confidence intervals use the percentile method from 5{,}000 cluster bootstrap resamples of journal-year cells (seed 20260322). ``Pre-DA-RT'' and ``Post-DA-RT'' use journal-specific cutoffs: APSR $\geq$2016, AJPS $\geq$2015, JOP $\geq$2015. Bootstrap intervals are wider for individual journals (5--11 clusters per period) but closely match the analytic intervals for pooled estimates.}
\end{minipage}
\end{minipage}

\clearpage

\subsection{Diagnostic Classification}\label{sec:failure}

Under the strict coefficient-level 100\% threshold (Section~\ref{sec:corpus}), \dnum{61} of the \dnum{254} papers with accessible materials fall below the fully-reproducible classification. Table~\ref{tab:gap-summary} summarizes their distribution across verdict categories: the majority (\dnum{41}, \dnum{67.2\%}) are largely reproducible---one or two unmatched coefficients out of dozens---while \dnum{20} (\dnum{32.8\%}) exhibit more substantial gaps. Tables~\ref{tab:diagnostic-gap} and~\ref{tab:diagnostic-repaired} below present hand-curated descriptions of \dnum{25} representative cases identified during deeper investigation. Table~\ref{tab:diagnostic-gap} lists \dnum{15} papers whose gaps remain unresolved---due to missing author data, pipeline limitations, or ambiguous code-to-table mappings. Table~\ref{tab:diagnostic-repaired} lists \dnum{10} papers where the workflow's adaptive execution repaired author-side packaging errors (missing dependencies, deprecated syntax, broken paths), recovering them to fully or near-fully reproducible status; the recovery verdict reflects the model-level criterion used during the diagnostic pass, and some of these papers fall just below the strict coefficient-level threshold in the final classification.

\begin{table}[!ht]
\centering
\caption{Distribution of Below-Fully-Reproducible Papers}
\label{tab:gap-summary}\small
\begin{tabular}{lcl}
\hline\hline
Verdict & N & Definition (coefficient-level match rate) \\
\hline
Largely reproducible    & 40 & $>80\%$ but $<100\%$ of coefficients match \\
Partially reproducible  & 11 & $50\%$ to $80\%$ of coefficients match \\
Not reproducible        & 10 & $<50\%$ of coefficients match \\
\hline
\textbf{Total}          & \textbf{61} & Papers with accessible materials, below fully \\
\hline\hline
\end{tabular}
\begin{minipage}{0.95\textwidth}
\vspace{4pt}
{\footnotesize \textit{Notes.} Of the \dnum{384} sampled papers, \dnum{130} are excluded from this table because the replication package is unavailable (\dnum{106}) or restricted (\dnum{24}). The \dnum{61} papers above are those that executed (or attempted to execute) but did not reach the strict \dnum{100\%} coefficient-match threshold required for \textit{fully reproducible}. Tables~\ref{tab:diagnostic-gap} and~\ref{tab:diagnostic-repaired} below provide hand-curated descriptions of selected representative cases drawn from this set.}
\end{minipage}
\end{table}

{\small
\begin{longtable}{@{}p{1.6cm}p{1.5cm}p{2.6cm}p{8.8cm}@{}}
\caption{Papers Remaining Below Fully Reproducible}
\label{tab:diagnostic-gap}\\
\toprule
\textbf{Paper} & \textbf{Verdict} & \textbf{Category} & \textbf{Why not fully reproducible} \\
\midrule
\endfirsthead

\multicolumn{4}{@{}l}{\tablename~\thetable{} \textit{(continued)}} \\
\toprule
\textbf{Paper} & \textbf{Verdict} & \textbf{Category} & \textbf{Why not fully reproducible} \\
\midrule
\endhead

\midrule
\multicolumn{4}{r@{}}{\textit{Continued on next page}} \\
\endfoot

\bottomrule
\endlastfoot

\multicolumn{4}{@{}l}{\textit{Author-caused unresolved gaps}} \\
\addlinespace[2pt]
Paper 1  & Partial    & Missing author data      & A main-text table requires a data file not included in the released package; only appendix or intermediate files are provided. \\
Paper 2  & Partial    & Restricted author data   & The README states that the full analysis depends on restricted registry data; only a subset of results is reproducible from released summary statistics. \\
Paper 3  & Partial    & Missing data object      & A main-text table depends on a data object not shipped in the public package; the released file contains only pre-computed summaries. \\
Paper 4  & Partial    & Missing code/spec.       & The released package exposes a richer model specification but not the code needed to reproduce a reduced model in the main table. \\
Paper 5  & Partial    & Missing author data      & The package omits a survey source file required for one of the main-text tables. \\
Paper 6  & Partial    & Paper/code divergence    & The shipped code runs and produces model output, but the remaining gap is between paper-reported values and the code-generated coefficients. \\
Paper 7  & Not repl.  & Missing data inputs      & One set of main-text tables replicates, but remaining tables require cleaned input files absent from the released package. \\
Paper 8  & Restricted & Proprietary data         & Key variables were purchased from a commercial provider and excluded from the public release, preventing construction of the central exposure measures. \\
Paper 9  & Restricted & Permissioned data        & The replication package requires variables from a government agency that are not included in the release. \\
Paper 10 & Largely    & Corrupt data file        & The main-text do-file aborts because a required data file is unreadable, blocking remaining table columns before estimation begins. \\
\addlinespace[4pt]

\multicolumn{4}{@{}l}{\textit{Pipeline-side deep semantic cases}} \\
\addlinespace[2pt]
Paper 11 & Not repl.  & Deep semantic recon.     & Raw models run and some transformed quantities are visible, but most remaining paper rows depend on ambiguous post-estimation transforms that cannot be exposed safely without deeper reconstruction. \\
Paper 12 & Partial    & Rendered-table recon.    & The missing table columns come from rendered output; recovered model families are not on the same displayed scale as the paper columns. \\
Paper 13 & Partial    & Incomplete rendered recovery & Printed regression-table output recovers some missing columns but does not provide a safe one-to-one reconstruction of the remaining table family. \\
Paper 14 & Largely    & Matching-pipeline recon. & The missing matched-model columns require a saved matched dataset and weighting artifact that are absent; the script does not complete the computationally intensive matching stage in batch. \\
\addlinespace[4pt]

\multicolumn{4}{@{}l}{\textit{Ambiguous cases}} \\
\addlinespace[2pt]
Paper 15 & Not repl.  & Appendix vs.\ main-table mapping & The available code prints an appendix matrix rather than the paper's main table, and there is no safe local proof that the appendix outputs correspond to the unmatched main-text rows. \\

\end{longtable}
}

\clearpage

{\small
\begin{longtable}{@{}p{1.6cm}p{1.5cm}p{2.6cm}p{8.8cm}@{}}
\caption{Author-Side Packaging Errors Repaired by the Workflow}
\label{tab:diagnostic-repaired}\\
\toprule
\textbf{Paper} & \textbf{Verdict} & \textbf{Category} & \textbf{What was repaired} \\
\midrule
\endfirsthead

\multicolumn{4}{@{}l}{\tablename~\thetable{} \textit{(continued)}} \\
\toprule
\textbf{Paper} & \textbf{Verdict} & \textbf{Category} & \textbf{What was repaired} \\
\midrule
\endhead

\bottomrule
\endlastfoot

Paper 16 & Fully   & Missing dependencies        & The package assumed user-written Stata dependencies without a working local setup; installing them recovered the core tables. \\
Paper 17 & Fully   & Fragile output layer        & Rendering commands stopped execution before results were fully exposed; bypassing those blockers recovered the paper. \\
Paper 18 & Fully   & Missing Stata dependency    & The released script called a user-written command without shipping or installing it; isolating the failing script and running the remaining regressions recovered the paper. \\
Paper 19 & Fully   & Missing Stata dependencies  & Legacy user-written commands were required for the last two main-text models; locating and installing them recovered the missing table columns. \\
Paper 20 & Fully   & Missing data \& deprecated output & The package referenced a missing data file and relied on deprecated output assumptions; isolating the broken script recovered the viable analysis chain. \\
Paper 21 & Fully   & Deprecated syntax           & Deprecated loop constructs and brittle setup paths blocked the main analysis pipeline; repairing those issues recovered the paper. \\
Paper 22 & Fully   & Missing Stata dependency    & The script depended on a user-written package for auxiliary output; bypassing the failing script preserved the substantive regression pipeline. \\
Paper 23 & Fully   & Missing sub-script          & The master do-file called a sub-script absent from the released package; running the surviving scripts directly recovered all main-text outputs. \\
Paper 24 & Fully   & Column-name bug             & The released analysis script expected a column name not present in the shipped data; correcting the mismatch recovered the main table models. \\
Paper 25 & Partial & Fragile RD bookkeeping      & Legacy post-estimation bookkeeping after valid RD estimates crashed the loop; stabilizing those blocks allowed most substantive models to run. \\

\end{longtable}
}

\clearpage

\section{System Architecture and Workflow}

This section describes the three-layer system architecture and details of the AI workflow.

\subsection{Three-Layer Architecture}

The AI workflow described in the main text, which combines deterministic orchestration with LLM-driven adaptive coordination, is implemented through a three-layer system. The layers are ordered by control flow: a LangGraph state machine governs deterministic routing, skill descriptions mediate task specification and accumulated failure-recovery knowledge consumed by the Adaptive Coordinator, and deterministic agent code executes all operations whose outputs must be numerically reproducible.

\paragraph*{Layer~1: LangGraph Orchestrator and Adaptive Coordinator.}

The orchestrator is a LangGraph state machine that manages the pipeline lifecycle. Each pipeline stage is represented as a node, and transitions between stages are encoded as explicit conditional edges that route tasks based on typed state attributes---for example, whether the previous stage succeeded, whether a validation gate flagged a fixable failure, or whether a fix loop has exhausted its retry budget. The routing layer is fully deterministic: for a given pipeline version and input state, the sequence of nodes traversed is reproducible. The \texttt{StateGraph} abstraction enforces a typed pipeline state and its built-in SQLite checkpointer persists state after every node, so the pipeline can resume from any point after a crash and every routing decision can be reconstructed from the checkpoint trace. When a stage fails in a manner the existing routers do not anticipate, the Adaptive Coordinator is dispatched to interpret the failure context, identify the root cause, and apply a minimal repair through tool-restricted file operations (\texttt{Read}, \texttt{Write}, \texttt{Edit}, \texttt{Bash}); control then returns to the graph and the deterministic pipeline re-executes. The Adaptive Coordinator does not perform statistical estimation, transform datasets, or modify numerical routines, so numerical results are produced exclusively by the deterministic execution layer.

\paragraph*{Layer~2: Skill descriptions and knowledge bases.}
Each agent is associated with a structured natural-language file (\texttt{SKILL.md}) that functions as both a formal interface specification and a persistent knowledge base.

The first component of the file defines the agent's contract: required inputs, expected outputs, permissible tools, execution constraints, and the sequence of subtasks. This specification ensures that each stage has a well-defined, inspectable contract that the LangGraph state machine and the Adaptive Coordinator can rely on.

The second component records accumulated failure patterns encountered during development and evaluation. Each entry documents the context in which a failure occurred, the root cause, and the generalized resolution rule. These entries are written in structured form to promote consistency across updates. When a new class of failure is resolved, the corresponding rule is added to the relevant skill file. This process expands the system’s coverage without modifying deterministic computation within a given pipeline version. Because skill files are version-controlled, each run is associated with a fixed and inspectable knowledge state.

\paragraph*{Layer~3: Deterministic agent code and diagnostic scripts.}
The bottom layer consists of deterministic program code that executes all file operations and statistical procedures. Each agent is implemented as an independent Python class responsible for a specific stage of the pipeline (e.g., metadata extraction, repository retrieval, specification parsing, code preparation, execution, or reporting). Agents operate only on explicit inputs and produce explicit outputs written to disk. They share no internal state.

Statistical estimation and diagnostic procedures are executed by explicit scripts in \texttt{R}, \texttt{Stata}, and \texttt{Python}. In particular, the full diagnostic suite is implemented in a standalone R script (\texttt{diagnostics\_core.R}) that consumes exported analysis datasets and produces structured diagnostic outputs. This script calls established statistical packages, including \texttt{estimatr} \citep{estimatr2024}, \texttt{fixest} \citep{berge2023}, \texttt{boot}, and \texttt{ivDiag} \citep{ivDiag2024}. Given the same inputs and the same pipeline version, this layer produces identical numerical results across runs.

\medskip

Information flows downward as instructions and dispatch decisions (Layer~1 to Layer~3) and upward as logs, intermediate artifacts, and error messages (Layer~3 to Layer~1). Adaptation occurs through controlled updates between runs: to skill descriptions in Layer~2, to deterministic agent code in Layer~3, or to the LangGraph routing layer in Layer~1 itself when a failure pattern crystallizes into a new conditional edge or gate router. Within a fixed version, numerical outputs depend exclusively on deterministic code and exported datasets.

This separation allows the system to remain adaptive in coverage while preserving computational determinacy in each execution. Human oversight operates at the boundary between Layers~1 and~2: proposed updates to knowledge bases or deterministic routines are reviewed before being committed. The result is an architecture that evolves across versions yet remains reproducible within version.

\clearpage

\subsection{Agents and Pipeline Stages}

The workflow is implemented through a set of specialized agents, each responsible for a distinct stage of the pipeline. The agents serve both the full-paper replication task (Phases~A and~B) and the design-specific diagnostic task (Phase~C). The IV diagnostic template is described separately in Section~\ref{sec:tests}.

\paragraph*{Phase~A: Acquisition and execution.} These tasks are distributed across four agents. The \emph{Profiler} extracts metadata and repository links from the published PDF, prioritizing URLs embedded in the paper over keyword-based search. The \emph{Librarian} retrieves the replication package from public repositories (Dataverse, GitHub, OSF, or direct links). The \emph{Janitor} prepares the code for automated execution: path dependencies are resolved, deprecated packages are handled, and environment assumptions are standardized. The \emph{Runner} executes the complete codebase and captures all estimation outputs---coefficients, standard errors, sample sizes, and clustering information---in structured form.

For the IV diagnostic task, the workflow additionally identifies the target 2SLS specifications in the code using language-specific parsing across Stata, R, and Python. All IV specifications reported in the paper's main-text regression tables are selected for diagnostic analysis. The analysis dataset for each specification is extracted and, when the original code is in Stata, independently re-estimated in R to verify cross-language consistency.

\paragraph*{Phase~B: Reproducibility verification.} The \emph{Extractor} extracts all coefficients reported in the paper’s regression tables using a vision-capable language model that reads rendered table images. The \emph{Matcher} compares extracted coefficients against code-generated outputs through precision-aware rounding and optimal one-to-one assignment, producing a per-table match rate and an overall reproducibility verdict. Phase~B applies only to the full-paper replication task.

\paragraph*{Phase~C: Diagnostic evaluation.} For papers that pass the reproducibility check and have a supported research design, the \emph{Skeptic} applies a standardized diagnostic template. For IV designs, this includes first-stage $F$-statistics, Anderson--Rubin tests, bootstrap confidence intervals, jackknife sensitivity, and OLS comparisons, implemented through established statistical packages \citep{ivDiag2024, estimatr2024, berge2023}. All diagnostic procedures are deterministic and version-controlled. The \emph{Journalist} compiles results into a standardized report with specification-level diagnostics and summary indicators.

\subsection{Adaptive Execution}

Replication packages span multiple programming languages, directory structures, and coding conventions. Many failure modes arise only when new materials are encountered. The workflow addresses this heterogeneity through a structured adaptation mechanism.

When the workflow encounters a recurring failure class---such as path mismatches, deprecated dependencies, or nonstandard syntax---the resolution is encoded as a generalized rule in the deterministic execution layer rather than as a paper-specific patch. Each resolution is documented in a structured knowledge base associated with the relevant pipeline stage, recording the failure context, root cause, fix, and scope of applicability. These knowledge bases are version-controlled: coverage expands across pipeline versions, while numerical behavior remains fixed within each version.

Adaptation proceeds through two complementary channels. Recurring failure classes that admit a deterministic rule are codified into the workflow itself at one of two levels: as a new Python rule inside the relevant stage agent (Layer~3), or as a new conditional edge or gate router inside the LangGraph state machine (Layer~1) so the failure pattern is dispatched automatically to the appropriate recovery node in subsequent runs. Patterns better captured as contextual guidance---rules of thumb, prioritization heuristics, documented edge cases---are recorded in the relevant skill description (Layer~2) and consumed by the Adaptive Coordinator during fix attempts. In all cases, updates occur between runs and are subject to human review before being committed. The empirical inventory of resolved issue classes is reported in Section~\ref{sec:inventory}.

\clearpage

\section{IV Diagnostic Template}
\label{sec:tests}

This section briefly summarizes the statistical diagnostics applied to each IV specification. The full motivation and interpretation are detailed in \citet{lal2024much}. The pipeline implements the same diagnostic template.

\paragraph*{Instrument strength.} Instrument strength is assessed using first-stage $F$-statistics. The effective $F$-statistic \citep{MontielOleaPflueger2013} serves as the primary indicator. Following common practice, $F < 10$ triggers a weak-instrument warning.

\paragraph*{Robust inference.} Inference robustness is evaluated using:

\begin{itemize}[leftmargin=2em]
\item The Anderson--Rubin (AR) test, which remains valid under weak instruments.
\item Bootstrap confidence intervals (including cluster bootstrap when clustering is used).
\item The $tF$ procedure \citep{Lee2022} in single-instrument cases, which adjusts critical values as a function of the first-stage $F$.
\end{itemize}

\paragraph*{Sensitivity analysis.} A leave-one-out jackknife procedure assesses the influence of individual observations or clusters on the IV estimate. Large changes relative to the baseline estimate trigger warnings.

\paragraph*{2SLS--OLS comparison.} An OLS model (without instruments) with the same outcome-treatment-covariates specification is estimated and compared to the 2SLS estimate.

\clearpage

\paragraph*{Original findings.} For comparison purposes, we reproduce Figure~5 of \citet{lal2024much}, which is open access, below.

\begin{figure}[!thbp]
\centering
\includegraphics[width=0.9\textwidth]{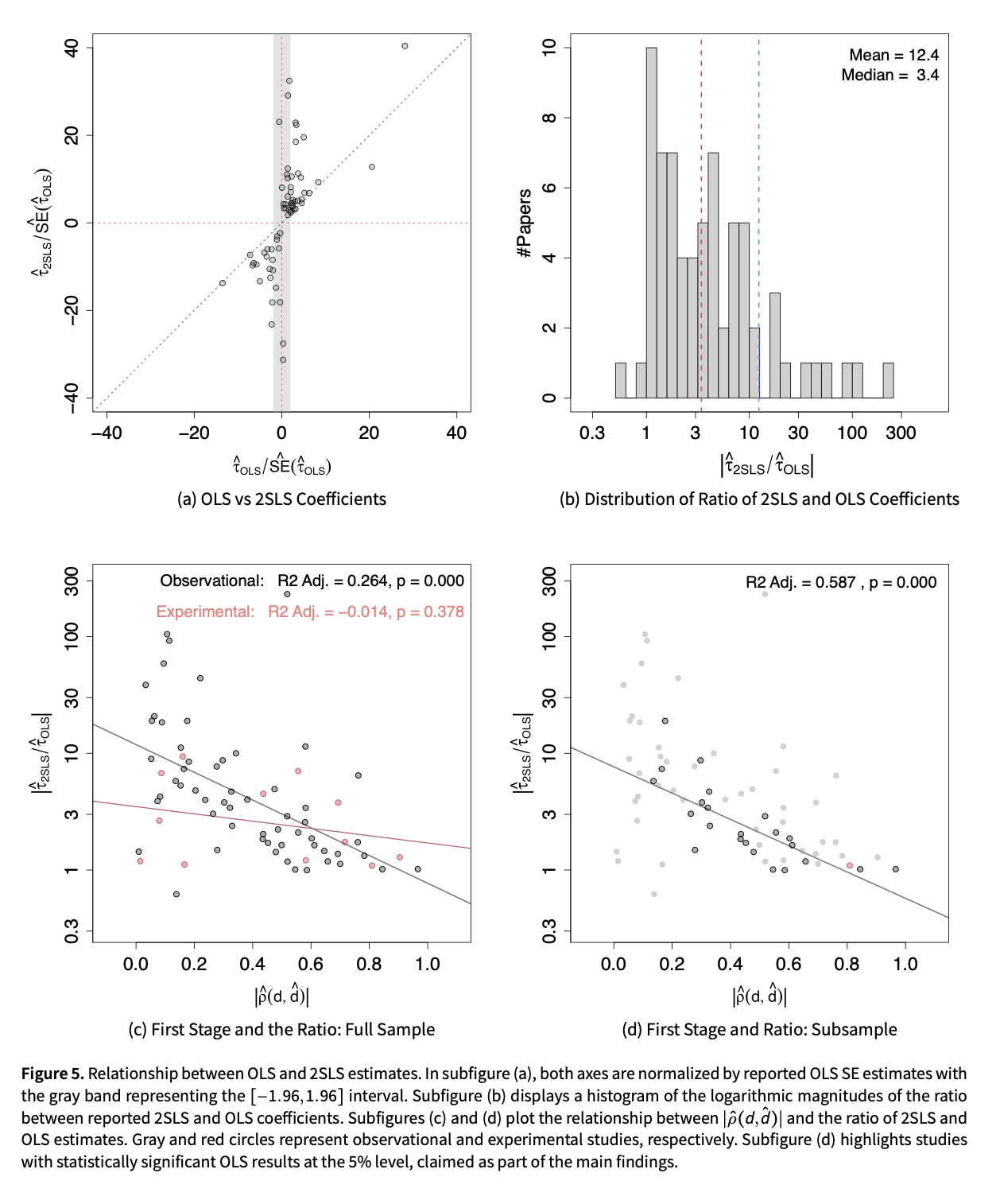}
\label{fig:fig5lal}
\end{figure}

\clearpage

\section{Empirical Inventory and Performance}
\label{sec:inventory}

This section documents the full set of implementation-level issues encountered and resolved during development and evaluation of the AI-assisted reproduction pipeline across \dnum{84} IV studies. Each entry records a distinct execution pattern, the agent(s) adjusted, and the corresponding resolution. The detailed inventory is organized into ten classes. Together, they provide a concrete account of the recurring irregularities that arise when replication packages are executed in a standardized, automated environment.

\subsection{Classes of Implementation Variation}

Across the \dnum{84} studies, we identified and resolved failures spanning ten broad classes. Table~\ref{tab:failure_summary} summarizes these classes, the affected pipeline stages, and representative failure patterns.

\begin{table}[htbp]
\centering
\caption{Classes of variation encountered and resolved across 84 IV studies}
\label{tab:failure_summary}
\footnotesize
\begin{tabular}{@{}p{5.3cm}cp{4.8cm}@{}}
\toprule
\textbf{Class} & \textbf{Stage(s)} & \textbf{Representative pattern} \\
\midrule
Path and environment & Janitor, Runner & Absolute paths, global macros, sub-file mismatches \\
Software and language & Profiler, Runner, Skeptic & Multi-language codebases, operator translation \\
Data format and encoding & Janitor, Runner, Skeptic & \texttt{.tab} ambiguity, factor encoding, quoting rules \\
Variable name mismatch & Skeptic, Profiler & Typos, prefix matches, unresolved macros \\
Stata syntax dialect & Janitor & \texttt{\#delimit} modes, merge syntax, \texttt{e(sample)} handling \\
Model specification structure & Skeptic, Profiler & Panel FE, subset conditions, bandwidth rules \\
Runtime resource constraints & Skeptic & Memory limits, jackknife timeouts \\
Graphics and interactivity & Janitor & Graphics devices, output table packages \\
Data acquisition & Librarian & Deprecated links, incorrect datasets \\
Code injection logic & Janitor & Delimiter mode, line-number drift \\
\bottomrule
\end{tabular}
\end{table}

These classes span the full pipeline: data acquisition, code parsing, environment preparation, execution, cross-language validation, and diagnostic computation. Some patterns reflect software dialect differences (e.g., Stata version changes); others reflect repository-specific formats or naming conventions. Many failures were silent and detectable only through cross-stage consistency checks.

The detailed inventory that follows lists each resolved issue, its manifestation, the deterministic repair implemented, and the responsible agent. The purpose is transparency rather than exhaustiveness of narrative: readers can trace each capability to a concrete failure pattern and code modification.

\subsection{Pipeline Performance}

Performance gains are substantial and empirically verifiable. The end-to-end success rate increased from approximately 63\% in the initial implementation to 92.5\% in the current version. These improvements were developed and evaluated on the same benchmark corpus; performance on previously unseen corpora may differ. Nevertheless, the resolved failure modes primarily reflect language-level and software-level conventions rather than idiosyncratic features of individual papers, suggesting that a meaningful portion of the gains should generalize.

The single largest improvement followed the adoption of a ``PDF-first retrieval architecture,'' in which the system first analyzes the published PDF to infer the likely location of the replication package before attempting data acquisition. This design substantially reduced incorrect dataset downloads and increased the success rate from 62.9\% to 88.6\%.

Approximately 40\% of resolved issue classes were first encountered during evaluation rather than anticipated ex ante. This pattern illustrates the limits of purely rule-based design. A fixed script can implement predefined rules, but it cannot account for patterns that were not foreseen. Here, new patterns were incorporated into the deterministic code base after diagnosis and review, expanding coverage across subsequent studies.

\subsection{Detailed Inventory of Resolved Issues}

Below we present a class-by-class inventory of resolved implementation issues.

\paragraph*{Class 1: Path and environment variability.}

Replication code is typically written for a specific directory structure and operating system. When executed in a standardized workspace, these assumptions often fail. The patterns below record the path and environment issues encountered and the corresponding repairs.

{\footnotesize
\begin{longtable}{@{}p{3.8cm}p{4.5cm}p{4.5cm}p{1.5cm}@{}}
\toprule
\textbf{Pattern} & \textbf{Manifestation} & \textbf{Resolution} & \textbf{Agent} \\
\midrule
\endfirsthead
\toprule
\textbf{Pattern} & \textbf{Manifestation} & \textbf{Resolution} & \textbf{Agent} \\
\midrule
\endhead
\bottomrule
\endfoot

Absolute paths &
\texttt{cd "C:\textbackslash Users\textbackslash john\textbackslash..."}, \texttt{setwd("/home/author/...")} &
Regex detection; replace with relative paths or comment out &
Janitor \\

Global path macros &
Stata \texttt{global datadir "C:\textbackslash..."} followed by \texttt{\$datadir} references &
Inline macro substitution before commenting out the \texttt{global} definition &
Janitor \\

\texttt{use} command variants &
Quoted paths, extensionless filenames, digit-prefixed names, macro-embedded paths &
Multiple regex patterns covering all observed variants &
Janitor \\

Sub-file reference mismatches &
\texttt{do script.do} when actual filename is \texttt{script\_rep.do} &
Fuzzy stem-matching: search for candidates when exact match fails &
Janitor \\

Subdirectory output files &
\texttt{analysis\_data.csv} generated in a subdirectory, not the expected root &
Recursive \texttt{rglob()} fallback search &
Runner \\

Platform-specific software paths &
macOS vs.\ Linux Stata installation paths &
Platform-aware path detection &
Runner \\

\end{longtable}
}

\paragraph*{Class 2: Software and language variability.}

The papers in the corpus use Stata, R, and Python, and several combine multiple languages within a single replication package. Supporting cross-language execution required systematic expansion of parsing and translation rules. The table below lists the software- and language-related issues resolved during development.

{\footnotesize
\begin{longtable}{@{}p{3.8cm}p{4.5cm}p{4.5cm}p{1.5cm}@{}}
\toprule
\textbf{Pattern} & \textbf{Manifestation} & \textbf{Resolution} & \textbf{Agent} \\
\midrule
\endfirsthead
\toprule
\textbf{Pattern} & \textbf{Manifestation} & \textbf{Resolution} & \textbf{Agent} \\
\midrule
\endhead
\bottomrule
\endfoot

Multi-language codebases &
Stata data preparation + R analysis in a single replication package &
Profiler parses \texttt{.do}, \texttt{.R}, and \texttt{.py} files simultaneously &
Profiler \\

Diverse IV commands &
\texttt{ivreg2}, \texttt{ivregress}, \texttt{reghdfe}, \texttt{ivprobit}, \texttt{ivtobit}, \texttt{rdrobust fuzzy}, manual 2SLS &
Expanding IV command pattern list; detecting manual two-stage implementations &
Profiler \\

Stata $\to$ R variable translation &
\texttt{l4.margin\_index2} becomes \texttt{l4margin\_index2} or \texttt{l4\_margin\_index2} in CSV &
Three-tier resolution: exact $\to$ dot-stripped $\to$ underscore-separated $\to$ recompute from base &
Runner, Skeptic, R core \\

Factor variable expansion &
\texttt{i.year} $\to$ dummy columns \texttt{\_Iyear\_2000}, \texttt{\_Iyear\_2001}, etc. &
Detect \texttt{i.} prefix; match expanded dummy patterns &
Runner, Skeptic \\

Time-series operators &
\texttt{L.}, \texttt{L4.}, \texttt{F2.}, \texttt{D.}, compound \texttt{L2D.var} &
Dual-side parsing chains in Python (Runner/Skeptic) and R (\texttt{diagnostics\_core.R}) &
Runner, Skeptic, R core \\

Truncated variable names &
Stata truncates to 32 characters with \texttt{\~{}} (e.g., \texttt{incumbvotesmajor\~{}t}) &
Prefix-plus-tilde pattern matching &
Runner \\

R formula objects with \texttt{update()} &
Base formula \texttt{f <- y \~{} x} modified by \texttt{update(f, . \~{} . + z)} &
Extract formula dictionary; apply \texttt{update()} rules to expand &
Profiler \\

R interaction shorthand &
\texttt{A * B} implying \texttt{A + B + A:B} &
Detect \texttt{*} patterns; add explicit \texttt{A:B} to control list &
Profiler \\

\end{longtable}
}

\paragraph*{Class 3: Data format and encoding variability.}

Replication materials are distributed in multiple data formats and encoding conventions. Differences in file types, string encodings, and export formats required explicit handling to ensure consistent downstream computation. The following entries summarize the issues encountered.

{\footnotesize
\begin{longtable}{@{}p{3.8cm}p{4.5cm}p{4.5cm}p{1.5cm}@{}}
\toprule
\textbf{Pattern} & \textbf{Manifestation} & \textbf{Resolution} & \textbf{Agent} \\
\midrule
\endfirsthead
\toprule
\textbf{Pattern} & \textbf{Manifestation} & \textbf{Resolution} & \textbf{Agent} \\
\midrule
\endhead
\bottomrule
\endfoot

\texttt{.tab} format ambiguity &
Dataverse stores as \texttt{.tab} (TSV); Stata expects \texttt{.dta} (binary) &
Detect \texttt{.tab} files; convert to CSV or rewrite \texttt{use} commands &
Janitor \\

Factor-encoded strings &
Column stored as \texttt{"0: Not low-education"} instead of numeric 0/1 &
Python preprocessing to convert to numeric before R diagnostics &
Skeptic \\

Lost variables in Dataverse export &
\texttt{pctcath} variable missing from \texttt{.tab} export &
Detect and flag as data-source limitation &
Skeptic \\

R backtick quoting &
Column \texttt{\_computed\_outcome} quoted as \texttt{`\_computed\_outcome`} by R internals &
Four-way lookup: raw $\to$ backtick $\to$ strip $\to$ fuzzy (9 call sites) &
R core \\

Destructive \texttt{gen} safeguard &
\texttt{capture drop X; gen X = expr} deletes \texttt{X} when \texttt{expr} fails &
Backup-restore pattern: rename $\to$ gen $\to$ restore on failure &
Janitor \\

Format conversion (\texttt{.dta} $\to$ CSV) &
Stata binary format unreadable by R/Python &
Automatic pandas-based conversion &
Runner \\

\end{longtable}
}

\paragraph*{Class 4: Variable name mismatches.}

Variable names extracted from code do not always match column names in the exported analysis dataset. Discrepancies arise from typos, truncation, macro expansion, encoding differences, or computed expressions. The patterns below record the matching rules added to reconcile these differences.

{\footnotesize
\begin{longtable}{@{}p{3.8cm}p{4.5cm}p{4.5cm}p{1.5cm}@{}}
\toprule
\textbf{Pattern} & \textbf{Manifestation} & \textbf{Resolution} & \textbf{Agent} \\
\midrule
\endfirsthead
\toprule
\textbf{Pattern} & \textbf{Manifestation} & \textbf{Resolution} & \textbf{Agent} \\
\midrule
\endhead
\bottomrule
\endfoot

Single-character typos &
\texttt{lcri\_euac1\_r} vs.\ \texttt{lcri\_euc1\_r} &
Four-tier resolution: exact $\to$ case-insensitive $\to$ Levenshtein $\leq 2$ $\to$ prefix match &
Skeptic \\

Missing suffixes &
\texttt{serfperc} vs.\ \texttt{serfperc1} &
Levenshtein distance matching &
Skeptic \\

Prefix matches &
\texttt{incumbvotes} vs.\ \texttt{incumbvotesmajorpercent} &
Bidirectional prefix detection &
Skeptic \\

Computed expressions &
\texttt{zero1(infeels-outfeels)} is a function expression, not a column name &
Expression detection + precomputation (min-max normalize, log transform) &
Skeptic \\

Unresolved Stata macros &
Metadata contains \texttt{\$controls\_z2} or \texttt{`controls\_z2'} literally &
Scan \texttt{command} field for \texttt{\$xxx} references; expand using macro dictionary &
Profiler \\

Unsplit cluster variables &
\texttt{"ccode year"} $\to$ R's \texttt{make.names()} produces \texttt{"ccode.year"} &
Space-based splitting normalization for cluster variables &
Skeptic \\

Unresolved macro passthrough &
\texttt{\$Z}, \texttt{\$C} passed to R as literal strings &
Guard: detect unresolved \texttt{\$macro} patterns before invoking R &
Skeptic \\

Case inconsistencies &
Variable names differ only in capitalization &
Case-insensitive matching (second tier of resolution chain) &
Skeptic \\

\end{longtable}
}

\paragraph*{Class 5: Stata syntax variants.}

Stata syntax differs across versions and programming styles. The pipeline encountered variations in delimiter modes, legacy commands, macro conventions, and wrapper structures. The following entries summarize the syntax-related adjustments implemented in the Janitor and related agents.

{\footnotesize
\begin{longtable}{@{}p{3.8cm}p{4.5cm}p{4.5cm}p{1.5cm}@{}}
\toprule
\textbf{Pattern} & \textbf{Manifestation} & \textbf{Resolution} & \textbf{Agent} \\
\midrule
\endfirsthead
\toprule
\textbf{Pattern} & \textbf{Manifestation} & \textbf{Resolution} & \textbf{Agent} \\
\midrule
\endhead
\bottomrule
\endfoot

\texttt{\#delimit ;} mode &
Semicolons replace newlines as statement terminators; multi-line commands &
Delimiter state tracking; treat content between semicolons as one statement &
Janitor \\

\texttt{\#delimit} abbreviations &
\texttt{\#d}, \texttt{\#delim}, \texttt{\#delimi} are all valid &
Extended regex: \texttt{\#d(?:e(?:l(?:i(?:m(?:i(?:t)?)?)?)?)?)?} &
Janitor \\

Old \texttt{merge} syntax &
\texttt{merge cow year using "file.dta"} (pre-Stata 11) &
Auto-detection and conversion to \texttt{merge m:1 ...} &
Janitor \\

Weight specifications &
\texttt{[aweight=w]}, \texttt{[pweight=w]} embedded in IV commands &
Separate weight parsing before variable-list extraction &
Runner \\

\texttt{e(sample)} filtering &
\texttt{ivreg2 ... if e(sample)} references prior estimation's sample &
Full-panel export with \texttt{janitor\_esample} flag column; R filters post-lag-computation &
Janitor, R core \\

\texttt{capture noisily} side effects &
Wrapping failed commands corrupts \texttt{e(sample)} and \texttt{e(b)} &
Comment out non-target IV commands entirely instead of wrapping &
Janitor \\

\texttt{e(sample)} restoration &
After \texttt{capture noisily} failure, \texttt{e(sample)} is invalid &
Backward scan to find estimation command that set \texttt{e(sample)}; inject re-estimation &
Janitor \\

User-defined commands &
\texttt{edvreg} as custom \texttt{ivreg2} wrapper &
Maintain list of known user-defined commands; detect \texttt{program define} blocks &
Janitor \\

Wrapper commands &
\texttt{parmby "ivreg2 ...", ...}, \texttt{bootstrap}, \texttt{jackknife} &
Wrapper exclusion list: preserve IV commands inside wrappers from commenting &
Janitor \\

Multi-line command end detection &
In \texttt{\#delimit ;} mode, commands span multiple lines until \texttt{;} &
Extended \texttt{\_find\_end\_of\_stata\_cmd()} to scan forward to semicolon in delimiter mode &
Janitor \\

\end{longtable}
}

\paragraph*{Class 6: Statistical model specification variability.}

The IV designs in the corpus differ in fixed effects, sub-sample conditions, clustering structures, weights, and model types. Supporting these variations required additional parsing and normalization rules. The table below lists the specification-related issues resolved.

{\footnotesize
\begin{longtable}{@{}p{3.8cm}p{4.5cm}p{4.5cm}p{1.5cm}@{}}
\toprule
\textbf{Pattern} & \textbf{Manifestation} & \textbf{Resolution} & \textbf{Agent} \\
\midrule
\endfirsthead
\toprule
\textbf{Pattern} & \textbf{Manifestation} & \textbf{Resolution} & \textbf{Agent} \\
\midrule
\endhead
\bottomrule
\endfoot

Panel fixed effects &
\texttt{xtivreg2, fe} requires \texttt{xtset panel time} &
Detect \texttt{xtset} in metadata; add FE columns to IV formula &
Skeptic \\

Stata \texttt{if} conditions &
\texttt{if year >= 2000 \& region == "east"} &
Extract condition; pass through to R \texttt{-{}-if\_condition} parameter &
Skeptic \\

R inline subsetting &
\texttt{data = df[df\$var == 0, ]} &
Manual addition of \texttt{subset()} to metadata \texttt{data\_prep} &
Skeptic \\

RDD bandwidth subsetting &
\texttt{rdrobust} computes bandwidth at runtime &
Compute bandwidth from execution log; hardcode as numeric subset condition &
Skeptic \\

Endogenous variable interactions &
\texttt{log(providers) * as.factor(year)} &
Identified as structural limitation; Journalist flags as ``Non-Linear Model Approximation'' &
Journalist \\

Non-linear IV models &
\texttt{ivprobit}, \texttt{ivtobit} &
Journalist generates information box noting linear approximation &
Journalist \\

Weighted regression &
Stata \texttt{[pweight=w]}, R \texttt{weights = sample.size} &
Cross-software weights extraction with fallback regardless of language &
Skeptic \\

Multi-equation output &
\texttt{ivprobit}/\texttt{ivtobit} produce two-line headers &
Multi-line header detection and merging helper functions &
Runner \\

\end{longtable}
}

\paragraph*{Class 7: Runtime resource constraints.}

The datasets in the corpus range from small cross-sections to large panels exceeding one million observations. These differences create variation in memory use and runtime. The following entries summarize the constraints encountered and the safeguards implemented.

{\footnotesize
\begin{longtable}{@{}p{3.8cm}p{4.5cm}p{4.5cm}p{1.5cm}@{}}
\toprule
\textbf{Pattern} & \textbf{Manifestation} & \textbf{Resolution} & \textbf{Agent} \\
\midrule
\endfirsthead
\toprule
\textbf{Pattern} & \textbf{Manifestation} & \textbf{Resolution} & \textbf{Agent} \\
\midrule
\endhead
\bottomrule
\endfoot

Large-dataset timeout/OOM &
1.26M rows (Ritter 2016); 115K rows (Lelkes 2017) &
\texttt{MAX\_OBS = 100{,}000} sampling cap with fixed seed for reproducibility &
R core \\

Cluster jackknife memory exhaustion &
Large cluster count causes OOM &
Fallback to observation-level jackknife (200 observations) &
Skeptic \\

Code execution timeout &
Some scripts exceed 600s &
Configurable hard timeout (\texttt{-{}-timeout}) &
Runner \\

Pre-computed lag column NAs &
Full-panel export has extra NAs in lag columns outside estimation sample &
Compare NA counts: if pre-computed has more NAs than base, recompute from base &
R core \\

\end{longtable}
}

\paragraph*{Class 8: Graphics and interactive commands.}

Many replication packages assume an interactive environment with an available graphics device and user input. In a batch execution setting, such commands cause interruptions or failures. The table below records the patterns identified and the corresponding handling rules.

{\footnotesize
\begin{longtable}{@{}p{3.8cm}p{4.5cm}p{4.5cm}p{1.5cm}@{}}
\toprule
\textbf{Pattern} & \textbf{Manifestation} & \textbf{Resolution} & \textbf{Agent} \\
\midrule
\endfirsthead
\toprule
\textbf{Pattern} & \textbf{Manifestation} & \textbf{Resolution} & \textbf{Agent} \\
\midrule
\endhead
\bottomrule
\endfoot

Common graphics commands &
\texttt{graph twoway}, \texttt{histogram}, \texttt{plot()}, \texttt{ggplot()}, \texttt{plt.show()} &
Regex-based commenting &
Janitor \\

Rare Stata graphics &
\texttt{cibplot}, \texttt{marginsplot}, \texttt{binscatter}, \texttt{spmap} &
Expanded graphics command list &
Janitor \\

Interactive commands &
\texttt{pause}, \texttt{View()}, \texttt{browser()}, \texttt{input()}, \texttt{breakpoint()} &
Comment out all interactive commands &
Janitor \\

Output table packages &
\texttt{modelsummary()}, \texttt{stargazer()}, \texttt{texreg()} &
Comment out (unnecessary for data extraction; may fail in batch mode) &
Janitor \\

Over-commenting false positives &
Function named \texttt{estimate\_plot\_data} incorrectly flagged as graphics &
Match complete function call patterns, not substrings &
Janitor \\

\end{longtable}
}

\paragraph*{Class 9: Data acquisition variability.}

Replication materials are hosted on multiple platforms with different API formats, URL conventions, and availability guarantees. Supporting these platforms required explicit handling of retrieval formats and error cases. The following entries document the acquisition-related issues encountered.

{\footnotesize
\begin{longtable}{@{}p{3.8cm}p{4.5cm}p{4.5cm}p{1.5cm}@{}}
\toprule
\textbf{Pattern} & \textbf{Manifestation} & \textbf{Resolution} & \textbf{Agent} \\
\midrule
\endfirsthead
\toprule
\textbf{Pattern} & \textbf{Manifestation} & \textbf{Resolution} & \textbf{Agent} \\
\midrule
\endhead
\bottomrule
\endfoot

Multiple hosting platforms &
Harvard Dataverse, GitHub, OSF, journal websites &
Multi-platform API support &
Librarian \\

Wrong dataset retrieval &
Keyword search returns unrelated datasets with similar titles &
PDF-first architecture: prioritize URLs from paper over search &
Librarian \\

API format errors &
Trailing slash in Dataverse API URL causes 404 &
Correct API URL formatting &
Librarian \\

Deprecated R packages &
\texttt{rgdal}, \texttt{rgeos}, \texttt{maptools} retired from CRAN in 2023 &
Comment out \texttt{library()} calls for known deprecated packages &
Janitor \\

Incompatible R packages &
\texttt{ri} package incompatible with R 4.x &
Flagged as unfixable &
--- \\

Expired repository URLs &
Repository taken offline or never publicly released &
Multi-tier retrieval with manual-supply fallback &
Librarian \\

\end{longtable}
}

\paragraph*{Class 10: Code injection logic variability.}

To extract analysis datasets, the Janitor inserts export commands into the original scripts. The correct insertion depends on delimiter modes, multi-line commands, and surrounding control flow. The entries below summarize the injection-related patterns resolved during development.

{\footnotesize
\begin{longtable}{@{}p{3.8cm}p{4.5cm}p{4.5cm}p{1.5cm}@{}}
\toprule
\textbf{Pattern} & \textbf{Manifestation} & \textbf{Resolution} & \textbf{Agent} \\
\midrule
\endfirsthead
\toprule
\textbf{Pattern} & \textbf{Manifestation} & \textbf{Resolution} & \textbf{Agent} \\
\midrule
\endhead
\bottomrule
\endfoot

Injection in \texttt{\#delimit ;} regions &
Injected code uses newline terminators; surrounding code uses semicolons &
Detect active delimiter mode; wrap injection with \texttt{\#delimit cr}/\texttt{\#delimit ;} &
Janitor \\

Multi-line command truncation &
Export block inserted between lines of a multi-line command &
Scan forward to complete command end before inserting &
Janitor \\

Line-number drift &
Insertions shift all subsequent line numbers; index-based tracking breaks &
Content-based detection (backward scan from marker) instead of index tracking &
Janitor \\

Full-panel vs.\ \texttt{e(sample)} export &
Some specifications need full panel for lag computation; others need only the estimation sample &
Detect \texttt{esample\_mode}; choose export strategy accordingly &
Janitor \\

\texttt{parmest} block handling &
Code between \texttt{parmest} and \texttt{restore} is interdependent &
Comment out entire block as a unit &
Janitor \\

\end{longtable}
}

\end{document}